\documentclass{aa}  

\usepackage[colorlinks=true,linkcolor=blue,citecolor=blue,filecolor=blue,urlcolor=blue]{hyperref}

\usepackage[]{natbib}
\usepackage{rotating}
\usepackage{orcidlink}
\usepackage{graphicx}
\usepackage{multirow}
\usepackage{float}
\usepackage{adjustbox}
\usepackage{placeins}

\usepackage{txfonts}
\usepackage{lscape}
\newcommand{\teff}{$T_{\rm eff}$} 
\newcommand{\logg}{$\log{g}$} 
\newcommand{\meta}{[Fe/H]} 
\usepackage{ulem}

\begin{document}

   \title{Stellar characterization with photometric colors from J-PLUS and 2MASS surveys}\titlerunning{Stellar characterization with J-PLUS and 2MASS surveys}

   \author{J. F. Aguilar\inst{\ref{UAM}, \ref{UMNG}}\orcidlink{0000-0002-4752-2784}
          \and
          P. Cruz\inst{\ref{CAB}}\orcidlink{0000-0003-1793-200X}
          \and
          E. Solano\inst{\ref{CAB}}
          \and
          P. R. T. Coelho\inst{\ref{USP}}
        \and A.~Ederoclite \inst{\ref{CEFCA},\ref{UA}}
        \and V. M.\ Placco \inst{\ref{NSF}}\orcidlink{0000-0003-4479-1265}
        \and P.~Mas-Buitrago \inst{\ref{CAB}}
        \and A. Alvarez-Candal \inst{\ref{IAA}}\orcidlink{0000-0002-5045-9675}
        \and F. Jiménez-Esteban\inst{\ref{CAB}}
        \and A.~J.~Cenarro\inst{\ref{CEFCA},\ref{UA}}
        \and D.~Crist\'obal-Hornillos\inst{\ref{CEFCA}}
        \and C.~Hern\'andez-Monteagudo\inst{\ref{IAC},\ref{ULL}}
        \and C.~L\'opez-Sanjuan\inst{\ref{CEFCA},\ref{UA}}
        \and A.~Mar\'{\i}n-Franch\inst{\ref{CEFCA},\ref{UA}}
        \and M.~Moles\inst{\ref{CEFCA}}
        \and J.~Varela\inst{\ref{CEFCA}}
        \and H.~V\'azquez Rami\'o\inst{\ref{CEFCA},\ref{UA}}
        \and J.~Alcaniz\inst{\ref{ON}}
        \and R.~A.~Dupke\inst{\ref{ON},\ref{MU}}
        \and L.~Sodr\'e Jr.\inst{\ref{USP}}
        \and R.~E.~Angulo\inst{\ref{DIPC},\ref{ikerbasque}}
          }

\institute{
    PhD Programme in Astrophysics, Doctoral School, Universidad Aut\'onoma de Madrid, Ciudad Universitaria de Cantoblanco, 28049 Madrid, Spain\label{UAM}
    \and
        Departamento de Matem\'aticas, Universidad Militar Nueva Granada, kil\'ometro 2 v\'ia Cajic\'a - Zipaquir\'a, Colombia, c\'odigo postal 110111. \label{UMNG}
    \email{john.aguilar@unimilitar.edu.co}
    \and
    Centro de Astrobiolog\'{\i}a (CAB), CSIC-INTA, Camino Bajo del Castillo s/n, E-28692, Villanueva de la Ca\~{n}ada, Madrid, Spain \label{CAB}
     \and Universidade de S\~ao Paulo, Instituto de Astronomia, Geof\'{\i}sica e Ci\^encias Atmosf\'ericas, 05508-090 S\~ao Paulo, Brazil \label{USP}
\and Centro de Estudios de F\'{\i}sica del Cosmos de Arag\'on (CEFCA), Plaza San Juan 1, 44001 Teruel, Spain \label{CEFCA} 
\and Unidad Asociada CEFCA-IAA, CEFCA, Unidad Asociada al CSIC por el IAA y el IFCA, Plaza San Juan 1, 44001 Teruel, Spain\label{UA}
\and NSF NOIRLab, Tucson, AZ 85719, USA \label{NSF}
\and Instituto de Astrofísica de Andalucía – Consejo Superior de Investigaciones Científicas (IAA-CSIC), Glorieta de la Astronomía S/N,E-18008, Granada, Spain \label{IAA}
\and Instituto de Astrof\'{\i}sica de Canarias, La Laguna, 38205, Tenerife, Spain \label{IAC}
\and Departamento de Astrof\'{\i}sica, Universidad de La Laguna, 38206, Tenerife, Spain \label{ULL}
\and Observat\'orio Nacional - MCTI (ON), Rua Gal. Jos\'e Cristino 77, S\~ao Crist\'ov\~ao,
20921-400 Rio de Janeiro, Brazil \label{ON}
\and University of Michigan, Department of Astronomy, 1085 South University Ave., Ann
Arbor, MI 48109, USA \label{MU}
\and Donostia International Physics Centre (DIPC), Paseo Manuel de Lardizabal 4, 20018
Donostia-San Sebastián, Spain \label{DIPC}
\and IKERBASQUE, Basque Foundation for Science, 48013, Bilbao, Spain \label{ikerbasque}
}

   \date{Received XXX September 15, 1996; accepted August 27, 2026}

  \abstract
    {}
    {We aim at deriving stellar atmospheric parameters based on the photometric data from the Javalambre Photometric Local Universe Survey (J-PLUS) in addition to near-infrared photometry from the Two Micron All-Sky Survey (2MASS).} 
    {Our method consists of a semi-supervised machine learning approach based on the k-means method combined with a modified K-nearest neighbors algorithm. This method compares the observed photometry to a set of reference data to estimate the stellar effective temperature (\teff), surface gravity (\logg), and metallicity (\meta) of stars from J-PLUS Data Release 3 (DR3). For that, a set of 105 colors was constructed using extinction-corrected photometric data from J-PLUS DR3 and 2MASS.
    The reference dataset used was composed by the MESA Isochrones \& Stellar Tracks (MIST) models, the Large Sky Area Multi-Object Fiber Spectroscopic Telescope (LAMOST), and the Apache Point Observatory Galactic Evolution Experiment (APOGEE) surveys.}
    {We estimate \teff, \logg, and \meta, for approximately 5.6 million stars from J-PLUS DR3, along with their errors.
    Our results were in agreement with the spectroscopic estimates of LAMOST and APOGEE. 
    We also applied a dimension reduction method, seeking greater efficiency by reducing the computation time and minimizing the information needed to calculate the stellar parameters, resulting in a subset of 11 colors. From this approach, stellar parameters were obtained for approximately six million stars, increasing the number of studied objects. The results obtained were of the same order, despite having slightly higher uncertainties compared to those obtained with 105 colors.}
    {Our results demonstrate the potential of using a method built from machine learning algorithms that does not require prior training. Additionally, we show that the proposed method allows the estimation of  reliable atmospheric parameters even when the available photometry did not fulfill all the photometric quality criteria. As part of our method, we define a neighborhood parameter, which assesses the reliability of our estimates and indicates that objects with smaller neighborhood values have lower uncertainties.}
    

   \maketitle
   \nolinenumbers

\section{Introduction}

The currently available information from large astronomical databases brings new possibilities for the study of stellar properties and evolution. For example, photometric surveys in the optical and near-infrared, such as the Sloan Digital Sky Survey \citep[SDSS;][]{2000AJ....120.1579Y}, the {\sl Gaia} mission \citep[{\sl Gaia};][]{GaiaMission}, the Two-Micron All Sky Survey \citep[2MASS;][]{skrutskie2006twoMASS}, The Southern Photometric Local Universe Survey \citep[S-PLUS;][]{2019MNRAS.489..241M}, and the Javalambre Photometric Local Universe Survey \citep[J-PLUS;][]{Cenarro2019}, have generated extensive accessible databases that enable statistical analyses and the application of machine learning (ML) methods.  
The ML techniques have made it possible to study specific characteristics along with a large number of associated variables. In particular, unsupervised and clustering techniques have been useful tools for automated classification. 
For example, the k-means clustering method \citep[]{jain1988algorithms} has been shown to be a robust approach, and it is regularly used in data mining and artificial intelligence to recognize patterns and trends \citep[e.g.,][]{everitt1995commentary, bishop2006pattern}. Among different applications in astronomy, such techniques are frequently applied to classify both galaxies and stars.  

Based on automated classification techniques, \citet[]{almeida2013automated} applied unsupervised k-means clustering to a massive set of stellar spectral data from SDSS, classifying stellar spectra into groups defined solely by the intrinsic properties of the data, distinguishing between dwarf and giant stars at similar temperatures. Using the same ML technique, \citet[]{morales2011systematic} systematically searched for extremely metal-poor galaxies in SDSS DR7 \citep[][]{2009ApJS..182..543A} using spectral shape analysis around H$\alpha$, and identified galaxies with metallicities ten times lower than solar. 
Other works used similar clustering techniques and spectroscopic information to distinguish among galaxies and quasars \citep[]{2008AIPC.1082.....B} or young stellar clusters with spectral information \citep[]{hojnacki2007x}.

There are various methodologies used to estimate stellar parameters, from classical spectroscopic approaches, recently combined with machine learning techniques, to methods based purely on photometry. 
\citet{2022AJ....163..152S} have derived effective temperatures (\teff), surface gravities (\logg), and metallicities (\meta) using a method based on convolutional neural network (CNN). For that, they used high-resolution spectra obtained from the Apache Point Observatory Galactic Evolution Experiment project \citep[APOGEE;][]{2013AJ....146...81Z} from the 17th data release \citep[DR17;][]{2022ApJS..259...35A}. The parameter determination was performed by comparing the observed spectra with a grid of synthetic spectra, such as those of \citet{1979ApJS...40....1K, 1993sssp.book.....K}, \citet{2005A&A...443..735C}, and \citet{2013A&A...553A...6H}.

Concerning photometry, \citet{2019A&A...622A.182W} developed a neural network-based pipeline Stellar Photometric Index Network Explorer (SPHINX), for the photometric estimation of stellar atmospheric parameters using data from the J-PLUS DR1 \citep{Cenarro2019}. The SPHINX system was trained with synthetic photometry derived from medium-resolution SDSS spectra and calibrated with the J-PLUS photometric system. 
Their method was applied to the globular cluster M15, retrieving metal-poor stars with \meta\ $<$ $-$2.0\,dex. 
\citet{Yang2022} used data from J-PLUS DR1, and from {\sl Gaia} DR2 \citep{GaiaDR2_2018}, and spectroscopic labels from the Large Sky Area Multi-Object Fiber Spectroscopic Telescope \citep[LAMOST;][]{LAMOST_DENG2012} to obtain atmospheric parameters for a set of about two million stars. 
They designed and trained a set of cost-sensitive neural networks to estimate \teff, \logg, and \meta, $\alpha$-elements ([$\alpha$/Fe]), and four other elemental abundances. 

Subsequent analyses of J-PLUS Data Release 2 \citep[DR2;][]{2021A&A...654A..61L} involved the development and application of stellar parameters estimation based on ensemble methods (SPEEM pipeline by \citealt{2022A&A...657A..35G}), which utilized machine learning models trained to estimate stellar atmospheric parameters. This approach enabled the identification of 177 very metal-poor star candidates.  \citealt{2023MNRAS.522.3898Q} used J-PLUS DR2, combined with {\sl Gaia} DR3 astrometry, to characterize 28 high-velocity stars. 
They used a convolutional neural network (CNN) and an $\chi^{2}$ fitting of spectral energy distributions (SEDs) using the Virtual Observatory SED Analyzer \citep[VOSA;][]{2008A&A...492..277B} to derive stellar parameters and investigate the kinematics, dynamics, and possible origins of these stars.

\citet{Wang2022} developed a similar work, where they used a support vector regression (SVR) algorithm to derive \teff, \logg, and \meta\ for around 2.5 million stars in J-PLUS DR1. Later, \citet{Yang2024} studied the same stellar parameters for around five million stars, based on colors from J-PLUS DR3 \citep[][]{lopez2023j} and {\sl Gaia} EDR3 \citep{Gaia_collaboration2021}, using methods such as kernel principal component analysis (KPCA) and Bayesian methods.

Some supervised approaches, such as neural networks, require extensive training. Moreover, they rely on training datasets that may not uniformly sample the photometric feature space. As a result, predictions in poorly represented regions effectively involve extrapolation, which can lead to biased or spurious parameter estimates.
We adopted a semi-supervised machine-learning approach based on the k-means clustering technique combined with a modified K-nearest neighbors (KNN) regression method \citep{Cover1967}. This strategy restricts the estimation of stellar atmospheric parameters to local regions of the feature space that are supported by reference stars, thereby reducing the impact of extrapolations. 
We estimated \teff, \logg, and \meta\ using photometry from J-PLUS DR3, combined with the near-infrared counterpart from 2MASS. An additional aim was to assess the effect that photometric quality may have on the inferred parameters.

This work is presented as follows. In Sect.~\ref{selection}, we present the models and observational data used. The implementation of the proposed method is described in Sect.~\ref{MLmethodology}. In Sect.~\ref{Results}, we present our results, and in Sect.~\ref{discuss} our discussion. Our conclusions are presented in Sect.~\ref{concl}.

\section{Data}\label{selection} 

\subsection{Photometric data from J-PLUS DR3}\label{target_data}

The J-PLUS survey, operating in the Javalambre Astrophysical Observatory (OAJ), observes the sky with a collection of 12 broad-, intermediate-, and narrow-band optical filters -- $u$, $J0378$, $J0395$, $J0410$, $J0430$, $g$, $J0515$, $r$, $J0660$, $i$, $J0861$, and $z$ -- covering important spectral features such as, for instance, the Balmer break region, H$\delta$, Ca\,II H and K lines, the G band, and the Ca\,II triplet \citep{Cenarro2019}. In its third data release \citep[J-PLUS DR3;][]{lopez2023j}, it has over 47 million objects observed from a large area of around 3284\footnote{After masking, the area is $\sim$2881\,deg$^2$ \citep[see][for details]{lopez2023j}.}\,deg$^2$ of the sky.

J-PLUS DR3 provides criteria for identifying objects with a high probability of being stars. We adopted the \texttt{SExtractor}'s \texttt{CLASS\_STAR} and the Bayesian star-galaxy separation \texttt{prob\_class\_star}, both available in the J-PLUS catalog, as described in \citet{2019A&A...622A.177L}. Both quantities take values from 0 to 1, with 0 corresponding to a galaxy and 1 to a star. We selected objects with values greater than or equal to 0.85 in both cases. By choosing this number, we minimized the differences between the two parameters mentioned above in objects assigned as stars. After we applied both class criteria, it resulted in a selection of 7\,400\,214 objects.

We also used photometric data in the  $J$, $H$, and $K_s$ bands from the 2MASS survey \citep[][]{skrutskie2006twoMASS}. We crossmatched the selected objects to 2MASS using the Virtual Observatory tool \texttt{TOPCAT}\footnote{\texttt{TOPCAT} is an interactive Tool for OPerations on Catalogues And Tables, available at \url{https://www.star.bris.ac.uk/~mbt/topcat/}, and all further data arrangements were performed using \texttt{TOPCAT}.} \citep{Taylor05,taylor2011astrophysics}.
For this crossmatching, we used a 5\,arcsec radius, taking this as the maximum difference between the coordinates in both catalogs.
We kept only those objects that have magnitudes available for the complete set of 15 filters (12 from J-PLUS and 3 from 2MASS), which resulted in a total of 6\,246\,451 stars in our final sample. 

The extinction map from \citet{1998ApJ...500..525S} is available in J-PLUS DR3 catalogs\footnote{Available at jplus.MWExtinction table within the J-PLUS database.}, and it was adopted for reddening corrections in J-PLUS magnitudes. This map provides the line-of-sight $E(B - V)$ color excess integrated to infinity, which is nonetheless reasonably reliable for J-PLUS DR3, as the stellar sample predominantly lies at high Galactic latitudes. 
The 2MASS magnitudes were also corrected from interstellar extinction using the same extinction maps, where the $E(B - V)$ color excess was retrieved\footnote{The color excess was obtained using the SFDQuery class from the dustmaps Python package \citep[][]{2018JOSS....3..695G}, which provides access to the extinction values by \citet{1998ApJ...500..525S}.} and the extinction in each band was computed according to the relation $ A_{\lambda} = R_{\lambda} \times E(B - V) $, with $R_{\lambda}$ values from \citet{2013MNRAS.430.2188Y}.

One of the purposes of this study is to assess the effect that quality of photometric data has on the inferred parameters, that is, if the method gives good approximations for the derived stellar parameters, even if the quality of the photometry is somewhat compromised. For that, we divided our sample into two subsets. The high-quality photometry (HQP) subset contains objects whose photometry has specific quality flags depending on the survey. The "AAA"\footnote{Photometric quality flag “A” in a 2MASS band indicates that the corresponding magnitude measurement has an S/N greater than 10 and a measurement uncertainty less than $\sim$0.109 mag.} flag is associated with a good quality of the photometry of the three 2MASS filters, as well as \texttt{FLAG} and \texttt{MASK\_FLAG} with a value of $0$\footnote{Objects with this value have no reported photometric extraction or image masking problems, indicating clean detections without saturation, decay complications, proximity to bright stars or edges, and are not affected by known masked artifacts in the image data.} for each one of the 12 J-PLUS filters. 
The objects in the HQP subset must be sufficiently bright to minimize the impact of noise and instrumental uncertainties, which is why we required $r \leq 22\,$ mag. Moreover, all objects must have magnitude errors less than or equal to 5\% in all 15 photometric bands, resulting in a subset of 1\,922\,570 stars. The remaining objects, those that do not meet one or more of the quality criteria mentioned above, compose the somewhat compromised photometry (SCP) subset, with 4\,323\,881 stars.

\begin{table*}
    \caption{Range of stellar parameters covered by the different datasets considered.}
    \label{DescriptionData}
    \centering
    \begin{tabular}{lccccccc}
        \hline\hline
        Source & N & \multicolumn{2}{c}{\teff\ (K)} & \multicolumn{2}{c}{\logg\ (dex)} & \multicolumn{2}{c}{\meta\ (dex)} \\

        & & Min & Max & Min & Max & Min & Max \\
        \hline
                \sl Reference data & \multicolumn{2}{c}{} & \multicolumn{2}{c}{} & \multicolumn{2}{c}{} \\

        LAMOST & 10\,787 & 3744 & 11720 & 0.932 & 4.900 & $-$2.441 & 0.059 \\
        APOGEE & 4\,768 & 3186 & 6305 & 0.652 & 5.067 & $-$1.945 & 0.324 \\
        MIST & 5\,510 & 2212 & 168845 & $-$1.093 & 7.972 & $-$4.000 & 0.500 \\

                \sl Validation data & \multicolumn{2}{c}{} & \multicolumn{2}{c}{} & \multicolumn{2}{c}{} \\

        LAMOST & 206\,505 & 3750 & 13422 & 0.719 & 4.900 & $-$2.470 & 0.602 \\
        APOGEE & 13\,136 & 3186 & 10568 & 0.2037 & 5.151 & $-$1.945 & 0.426 \\

                \sl Comparison data & \multicolumn{2}{c}{} & \multicolumn{2}{c}{} & \multicolumn{2}{c}{} \\

        Reference 1 & 1\,079\,802 & 3578 & 8527 & 1.309 & 5.653 & $-$2.552 & $-$0.001 \\
        Reference 2 & 2\,385\,202 & 2537 & 7700 & 1.002 & 5.099 & $-$5.018 & 0 \\
        \hline
    \end{tabular}
    \tablebib{
(1)~\citet{Wang2022};
(2)~\citet{Yang2024}.
}
\end{table*}

\subsection{Reference data}\label{reference data}

We needed a reference sample with known \teff, \logg, and \meta\ values, and magnitudes available in all 15 bands to infer the stellar atmospheric parameters of the stars in our sample, as further described in Sect.~\ref{MLmethodology}. 
We gathered two different subsets: from spectroscopy and from stellar models.

The LAMOST survey \citep[][]{LAMOST_DENG2012} is conducted using a 4-meter quasi-meridian reflecting Schmidt telescope with 4000 fibers distributed over a field of view with 5\,deg diameter \citep{2012RAA....12.1197C}. The stellar parameters available in LAMOST DR9\footnote{Available at \url{http://www.lamost.org/dr9/v1.1/}.} \citep[][]{bai2021first} were estimated from low-resolution (R $\sim$ 2000) SDSS-like optical spectra using the LAMOST Stellar Parameter Pipeline \citep[LASPM;][]{2015RAA....15.1095L}.

We crossmatched our complete sample with LAMOST DR9, considering a 5\,arcsec search radius, to obtain the spectroscopic labels needed, that is, \teff, \logg, and \meta\ values derived from spectroscopy to be associated with the magnitudes in all 15 bands (J-PLUS and 2MASS). It is worth emphasizing that the spectra were not used in our method.
We then selected those objects that had LAMOST spectra with S/N greater than ten and errors in their LAMOST stellar parameters of less than 5\%, to ensure greater accuracy in the derived parameters, resulting in a total of 217\,292 objects. 
A pseudo-random sample was defined in order to reduce the calculation time but keeping a sufficient number of objects that represented the different stellar parameter values available in LAMOST, composing a subset with 10\,787 objects. Its coverage in stellar parameters is shown in Table~\ref{DescriptionData}. 

Similarly, we performed a crossmatch with APOGEE DR17 \citep{2022AJ....163..152S}, which was conducted using two spectrographs mounted at two 2.5-meter telescopes -- one at the Apache Point Observatory (APO), and one at the Las Campanas Observatory \citep[LCO;][]{1973ApOpt..12.1430B, 2006AJ....131.2332G}, with the ability to observe up to 300 sources in the H band (1.51–1.7 $\mu$m), with a resolution of R$\sim$22,500, a field of view of 3$^{\circ}$ diameter in APO, and 2$^{\circ}$ diameter in LCO \citep{2010SPIE.7735E..1CW, 2019PASP..131e5001W}. 
We selected those objects with an error in stellar parameters of less than 5\%, resulting in a total of 17\,904\footnote{These data are available at \citet{2022yCat..51630152S}.} objects. As in LAMOST, a subset of 4\,768 objects was selected through a pseudo-random sampling process, ensuring that the color distribution of the subset reproduces the one from the crossmatched sample. This representative subset was also used as reference data, as also shown in Table~\ref{DescriptionData}.

The APOGEE dataset was incorporated as reference data to cover ranges of atmospheric parameters that were not covered by the LAMOST reference data. It is worth emphasizing that in both cases the observed spectra were not used, only the derived stellar parameters were adopted as spectroscopic labels. 
The remaining LAMOST and APOGEE objects were considered validation data, as explained in Sect.~\ref{comparisondata}.

\begin{figure}
	\centering
	\includegraphics[width=0.95\columnwidth]{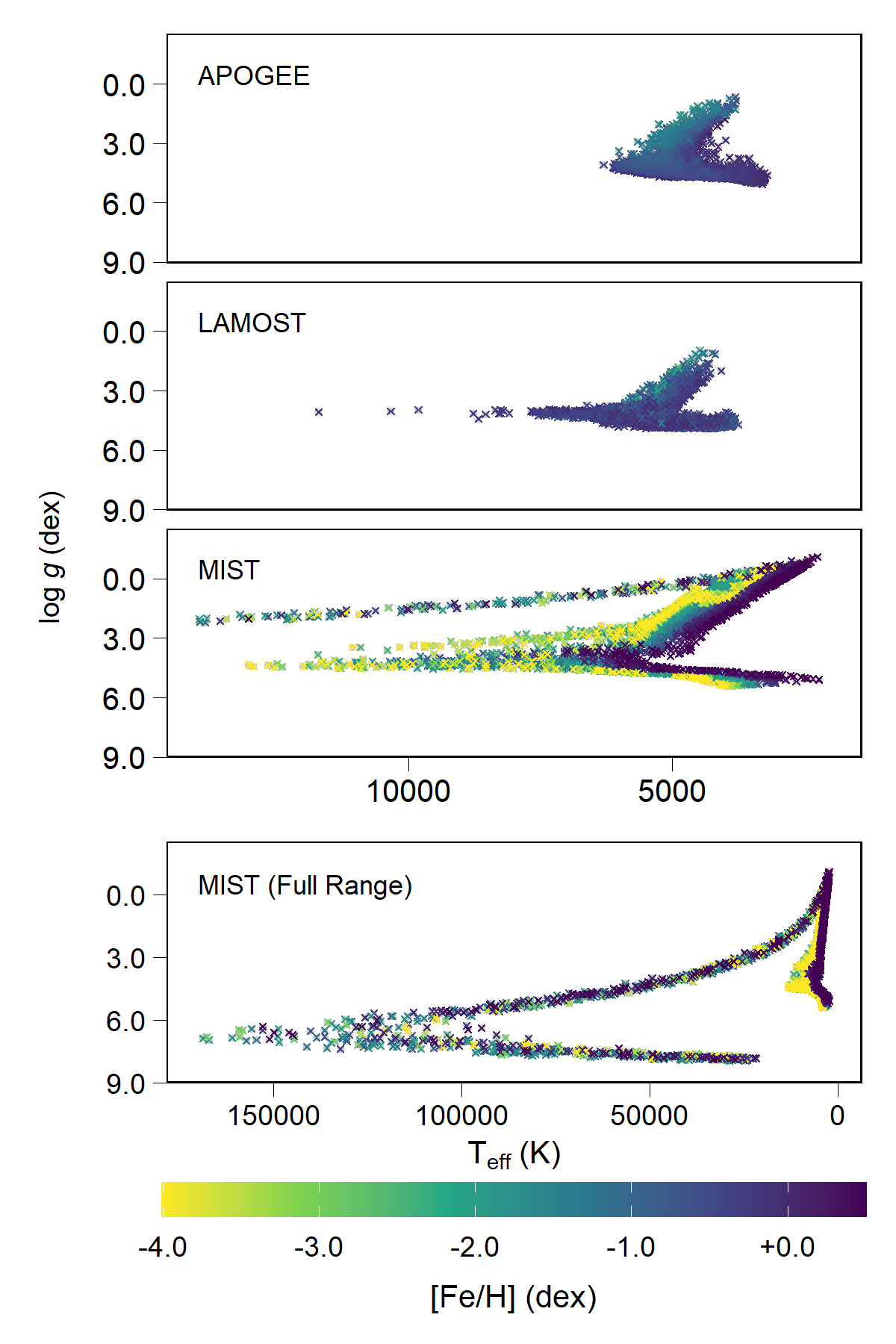}
	\caption{Kiel diagrams for the reference dataset, including APOGEE, LAMOST, and MIST (2 times), in order from top to bottom. The upper three panels display observed data from LAMOST and APOGEE, and theoretical models from MIST, limited to \teff\ between 2000 and 14000\,K. The bottom panel shows the full temperature range of the MIST models, from 2212 to 168845\,K.}
	\label{Kiel}
\end{figure}

Stellar models were adopted to expand the parameter's range covered by our reference sample. Synthetic photometric data and the associated stellar parameters were obtained from the MESA Isochrones and Stellar Tracks\footnote{Available at \url{https://waps.cfa.harvard.edu/MIST/}.} project \citep[MIST;][]{2016ApJS..222....8D, 2016ApJ...823..102C}, which is built on the Modules for Experiments in Stellar Astrophysics software \citep[MESA;][]{2011ApJS..192....3P, 2013ApJS..208....4P, 2015ApJS..220...15P}. 
From these models the stellar parameters -- \teff, \logg, metal abundance ($Z$) -- predicted by MESA are extracted and projected into the observational plane using synthetic photometry. 
The photometry is generated via bolometric correction tables computed from the C3K grid \citep{2016ApJ...823..102C}, which is based on the ATLAS12/SYNTHE model atmospheres \citep{1970SAOSR.309.....K, 1993sssp.book.....K}. Although MESA provides atmospheric parameters and evolutionary structure, the magnitudes in filters such as J‑PLUS and 2MASS are derived from Kurucz-based color-temperature relations and synthetic spectra, which means that the final results are based on Kurucz predictions processed using MESA outputs. By integrating both components, the MIST Packaged Model Grids\footnote{\url{https://waps.cfa.harvard.edu/MIST/model_grids.html}} give the photometry and the stellar parameters: synthetic photometry in J‑PLUS and 2MASS, together with the corresponding values of \teff, \logg\,, and $Z$, for stars with metallicities ranging from \meta \, $\approx$ –4.0 to +0.5\,dex.

The J-PLUS filter set is not available in the MIST database among the available integrated synthetic photometry; therefore, we extracted those generated for 
S-PLUS as it has a filter system which is almost identical\footnote{The S-PLUS filter system from the T80-South telescope and its large-format camera are also duplicates of the system installed at T80/JAST telescope used in J-PLUS \citep{2019MNRAS.489..241M}.} to the one from J-PLUS. Finally, our synthetic reference subset 
contains 5\,510 stellar models covering isochrones with $9\leq \log(age)\leq 10.1$ and v/vcrit=0.0. This range was chosen to avoid very young phases (< 1 Gyr), which include pre-main-sequence stars. The coverage of parameters is also shown in Table~\ref{DescriptionData}.  

The Kiel diagram, defined as the distribution of stars in the \logg – \teff\ plane, can be used as a diagnostic tool to assess the reliability of stellar parameters. Figure~\ref{Kiel} shows such a diagram, illustrating that the reference sample contains a broad range of stellar parameters. Although early-type or extremely metal-poor stars are the minority of objects expected in J-PLUS DR3, LAMOST, or APOGEE, they are represented in the MIST models adopted to allow a wider diversity of parameters and their corresponding photometry in the studied 15 bands. 
The MIST models are shown twice in Fig.~\ref{Kiel} to highlight the stellar parameter ranges similar to those obtained from spectroscopy and also to show the complete range covered by our reference dataset. 

\subsection{Validation and comparison data}\label{comparisondata}

We defined a sample of stars that have their stellar parameters estimated from different methodologies and that were not considered as a reference to assess the reliability of our results. 
They were divided into two groups: the validation set, with parameters estimated from spectroscopy, and the comparison set, with values estimated from photometric data and ML techniques. 

We selected 206\,505 objects from LAMOST DR9 as a validation set, those that were not included in the reference dataset (see Sect.~\ref{reference data}). Also, we added the remaining set of 13\,136 objects from APOGEE DR17. The stellar parameter coverage for the validation dataset (for both samples, separately) is presented in Table~\ref{DescriptionData}.

Previous works have applied ML techniques to infer stellar parameters from J-PLUS data. We selected two of them to compare our results with, which comprise our comparison sample. They are briefly summarized below.

\citet{Wang2022} estimated the same three stellar parameters for $\sim$2.5 million stars in J-PLUS DR1 using an SVR algorithm. The information used in training the data was the 12-band photometry from J-PLUS DR1, crossmatched with spectrum-based catalogs from LAMOST \citep[LAMOST DR7;][]{2022yCat.5156....0L} -- low resolution spectra (LRS) and medium resolution spectra (MRS)\footnote{\url{http://dr7.lamost.org/catalogue}} -- from APOGEE \citep[APOGEE DR16;][]{2020AJ....160..120J, 2020ApJS..249....3A}, and from the Sloan Extension for Galactic Understanding and Exploration \citep[SEGUE;][]{2009AJ....137.4377Y}, specifically the SEGUE stellar parameter pipeline \citep[SSPP;][]{2008AJ....136.2022L, 2008AJ....136.2050L}. They built a catalog with 2\,493\,424 stars, where the estimated root mean square errors for \teff, \logg, and \meta\ are 160\,K, 0.35\,dex, and 0.25\,dex, respectively. We performed a crossmatch between their catalog and our sample, excluding those that presented a quality flag\footnote{\citet{Wang2022} reported in their catalog, with an asterisk, some objects with less reliable stellar parameters.} or estimated errors over 5\% in the derived parameters, finding 1\,079\,802 objects in common. These objects were used as a comparison set, as shown in Table~\ref{DescriptionData}. 

\citet{Yang2024} adopted kernel principal component analysis (KPCA) and Bayesian methods to derive stellar parameters and elemental-abundance ratios for around five million stars (mostly dwarfs) based on stellar colors from J-PLUS DR3 and {\sl Gaia} EDR3. To build their model, they resorted to a large spectroscopic training set, which included catalogs such as SEGUE, LAMOST DR9, APOGEE DR17, and the GALactic Archaeology with HERMES \citep[GALAH DR3;][]{2021MNRAS.506..150B} surveys. These objects were filtered, selecting only those with a 5\% error on the stellar parameters, and a reliability flag\footnote{\citet{Yang2024} also adopted a quality flag to define the reliability of the estimated values and we adopted $flag_{[\mathrm{Fe/H}]} \geq 0.9$.}. The 2\,385\,202 objects in common with our sample were also taken as part of our comparison set (see Table~\ref{DescriptionData}).

\section{Methodology}\label{MLmethodology}

Just as objects with similar spectroscopic profiles have similar \teff, \logg, and \meta\ values, it is to be expected that the behavior of stellar photometric colors will also be similar. Validating this idea implies thinking that objects with this information, which is represented here by the 105 colors under study, should also be similar to each other and therefore, when represented in an $\mathbb{R}^{105}$ space, should be grouped together. 
Considering that photometric observations have their associated errors and that they may have limitations in the inference of stellar parameters, we defined the concept of neighborhoods as regions in the $\mathbb{R}^{105}$ space where objects with similar colors, which may be associated with similar stellar parameters, would be gathered. These neighborhoods may have different boundaries for different types of stars. 

We adopted a modified version of the procedure described in \citet{Garrido2019, Cruz2022}, based on the k-means algorithm \citep[]{jain1988algorithms}, to derive the stellar parameters of millions of stars in J-PLUS DR3. 
The k-means algorithm, an unsupervised method, is combined with a modified version of the K-nearest neighbors regression method \citep[KNN;][]{Cover1967}, a supervised approach in which the k neighbors that are contained within a given radius are taken instead of a typical constant number of neighbors.
The method was implemented in a set of 105 colors, generated from the combination of the 12 J-PLUS bands ($u$, $J0378$, $J0395$, $J0410$, $J0430$, $g$, $J0515$, $r$, $J0660$, $i$, $J0861$, $z$), with those of 2MASS ($J$, $H$, $K_s$). 
The method was developed using \texttt{R} software \citep{RSoftware}, which consists of five steps as described in detail below.

Taking into account colors and stellar parameters from the reference data set (Sect.~\ref{reference data}), we defined the number of clusters with similar characteristics that should be formed (step 1). To determine the optimal number of clusters, we applied the Hartigan test \citep{hartigan1975clustering}, which is iterated 50 times to see a trend\footnote{The differences in each of the iterations are due to the fact that the
centroids are taken randomly.} in the results and derive the optimum number of clusters. 
By analyzing the distribution of Hartigan statistics over multiple iterations, we ensured the stability of the clustering solution. The results consistently indicated a tendency to form 19 clusters, as determined from the median value of the Hartigan test across the 50 runs.

Once the number of groups was established, the k-means method was implemented as follows. Taking again the reference sample as input data, we distributed them into 19 clusters and defined the cluster centroids (step 2). 
This distribution is presented in Table~\ref{Kmeansphase1}, showing that both synthetic (from MIST) and observational (from LAMOST and APOGEE) data that share the same cluster have similar characteristics in some of their stellar parameters. 
Clusters 1, 2, 7, 11, and 17 contain spectroscopic reference objects together with MIST models, whereas the remaining clusters are populated only by MIST models and should therefore be interpreted as model-constrained regimes. In the clusters where synthetic and spectroscopic reference objects coexist, both types of data show broadly similar trends in the 105-dimensional color space, suggesting that the models and the spectroscopic samples occupy compatible regions of the color space. However, this does not imply that model-dominated regions have the same empirical support as those constrained by LAMOST and/or APOGEE. Estimates obtained in clusters populated only by MIST models should therefore be used with caution since their reliability depends primarily on the adequacy of the adopted synthetic photometry. Moreover, discrete sampling of the synthetic grid, particularly in \meta\ (with steps of 0.5\,dex), can produce wider gaps in the Kiel diagram than those observed in clusters that include spectroscopic reference objects. Another relevant consideration is that some clusters contain a larger fraction of objects with parameters derived from LAMOST than from APOGEE, or vice versa. Consequently, the relative representation of the two surveys within a given cluster can influence their respective contributions to the parameter estimates obtained for target sources assigned to that cluster.

In the second step, we also estimated pseudo-radii\footnote{They are called pseudo-radii because they make non-spherical shapes in the $\mathbb{R}^{105}$ space.} to describe a neighborhood of objects with similar colors, and therefore stellar parameters, inside every proposed cluster. Pseudo-radii (hereafter radii) are defined from the calculated distance (in $\mathbb{R}^{105}$) of each reference object to the centroid of its cluster. 
It is worth emphasizing that these distances generated a non-normal distribution, leading to radii that can take different values. Given this heterogeneity, we defined ten radii regions based on statistical measures for each cluster, corresponding to the minimum, the fifth, the 25th, the 75th, the 95th, and the 99th percentiles, the mean, the median, the mode, and the maximum values\footnote{The order of these values varies per cluster due to the heterogeneity of the distances described above.}. A number from one to ten (hereafter neighborhood) was associated with every region, with one being assigned to the smallest radius and ten to the largest one.

Because neighborhood radii are defined independently within each cluster, the same neighborhood index does not correspond to the same absolute radius across the different clusters. Consequently, two targets with the same neighborhood index need not be close to each other in the 105-dimensional color space. Having defined the clusters and their centroids, we implemented the k-means algorithm for the second time to include our objects of interest (step 3). For that, we calculated the Euclidean distance of each object in our target sample (Sect.~\ref{target_data}) in the $\mathbb{R}^{105}$ space to each of the 19 centroids defined in step 2. The targets were considered to be part of the clusters with the closest centroid. 

We then searched for the nearest reference objects within each assigned cluster to estimate the stellar parameters of the objects in the target sample (step 4). This search was performed independently for each target star, using the target position in the 105-dimensional color space as the center of the neighborhood. The neighboring reference objects were identified using the minimum neighborhood radius, from 1 to 10, which contained at least one reference object within the assigned cluster.

This procedure was implemented as follows. For a given target star, we first considered the radius associated with neighborhood 1, centered on the position of that target in the color space. If no reference object was found within this radius, the search region was progressively increased to neighborhood 2, and so on, until reference objects were found. If no reference object was found within any of the ten neighborhood radii, the stellar parameters were not estimated for that target object.

Finally, the three stellar parameters (\teff, \logg, \meta) were obtained for each target using a modified KNN method (step 5), which is based on the mean values considering neighboring reference data, that is, those within the defined radius. The associated standard deviations were taken as the corresponding errors for the estimated parameters. These uncertainties therefore quantify the local dispersion among the stellar parameter labels of the neighboring reference objects. They do not explicitly account for possible systematic offsets between LAMOST, APOGEE, and MIST.

\section{Results}\label{Results}

Our method was applied to more than six million objects in our sample (Sect.~\ref{target_data}). Among them, around 3\% had previously been spectroscopically characterized by LAMOST and APOGEE. In Sect.~\ref{AllColorsResults}, we describe the overall characteristics of our full dataset, while in Sect.~\ref{SpectroscopicContrast}, we compare our estimates to those surveys to assess the reliability of our results.

\subsection{Overview of results with all colors}\label{AllColorsResults}

By implementing our method, we inferred the three stellar atmospheric parameters for 5\,648\,486 out of the 6\,246\,451 J-PLUS DR3 objects with available photometry in all 15 bands (J-PLUS and 2MASS), using the full set of 105 colors. The method was unable to calculate the parameters for 597\,965 objects, which will be discussed later in this section.

As an example, the derived stellar parameters and their associated uncertainties for a small set in our sample are presented in Appendix~\ref{AppenA} (Table~\ref{105Results}), along with the corresponding photometric quality flag. A full version of the table is available online in VizieR.

As mentioned previously, the uncertainties were estimated by the standard deviation of the stellar parameters of neighboring reference objects. In cases where only one neighbor was available (i.e., only one comparison object fell within the defined neighborhood), the uncertainty was set to the average error of all objects with the same photometric quality and neighborhood.
Our results show  that stars with HQP (HQP subset) found at smaller neighborhood radii tend to exhibit smaller errors.

Table~\ref{Abstract105ColorsVsReferenceData} summarizes how nearly 5.6 million stars are grouped by their spectral types. The estimated \teff\ were associated with a spectral type according to the semiempirical relations by \citet{Pecaut13}\footnote{Available at \url{https://www.pas.rochester.edu/~emamajek/EEM_dwarf_UBVIJHK_colors_Teff.txt}.}. As shown in the table, the majority of the sample consists of G and K-type stars. 

\begin{table}
     \caption{Summary of our sample using 105 colors, according to their spectral types. $N$ is the number of objects in each class.}
    \label{Abstract105ColorsVsReferenceData}
    \centering
\begin{tabular}{lcccccc}
    \hline\hline
    \teff\ range & Spectral & N \\
    (K) & type & \\
    \hline
    10401--33300 & B &  1\,299    \\
    7401--10400   & A & 10\,568     \\
    6181--7400   & F & 256\,115    \\
    5326--6180   & G & 2\,368\,356  \\
    3891--5325   & K & 2\,471\,419  \\
    2310--3890   & M & 540\,729   \\
    \hline
\end{tabular}
\end{table}

Figure~\ref{Kiel2} shows the distribution of the obtained stellar parameters in a Kiel diagram for the stars in our sample, color-coded by their estimated metallicities and separated by their neighborhood radius (see Sect.~\ref{MLmethodology} for details). The distribution of results in neighborhoods, grouped by photometric quality, is summarized in Table~\ref{statsSample}. The adopted reference data (both observational and synthetic) is also shown in gray. 

It can be noted in Fig.~\ref{Kiel2} that for larger neighborhoods (five or higher; six bottom panels), there are objects to which the method assigned stellar parameters that were not contemplated in the reference data. The majority of these 2913 stars, which represents 0.05\% of the sample with derived stellar parameters, is part of the SCP subset (97\% of these objects), and the majority of them belong to neighborhoods 9 and 10. 
The adoption of large neighborhood radii generated associations with reference stars with heterogeneous parameters, generating solutions with combinations of \teff, \logg, and \meta\ not present in the reference set.
Adopting the standard deviation of neighboring reference parameters as the uncertainty associated with each estimate allows us to identify these particular cases. For the objects analyzed above, the estimated uncertainties range from 230 to 2309 K in \teff, from 0.75 to 1.53 dex in \logg, and from 0.50 to 2.03 dex in \meta. Together with their reported neighborhood values, these unusually large uncertainties provide further warning about the reliability of the corresponding parameter estimates. They are substantially larger than the uncertainties typically obtained with our method. Table \ref{MeanErrors105Colors} summarizes the general behavior of these quantities and shows that, in general, they increase as the neighborhood index, our indicator of reliability, increases.

\begin{table}
    \caption{Mean uncertainties in the stellar parameters estimated using 105 colours, as a function of the neighborhood index.}
    \label{MeanErrors105Colors}
    \centering
        \begin{tabular}{cccc}
        \hline\hline
        Neighbor- &
        $\bar{\sigma}_{T_{\rm eff,105}}$ &
        $\bar{\sigma}_{\log g_{105}}$ &
        $\bar{\sigma}_{{\rm [Fe/H]}_{105}}$ \\
        hood & (K) & (dex) & (dex) \\
        \hline
        1  & 92  & 0.133 & 0.089 \\
        2  & 81  & 0.071 & 0.118 \\
        3  & 103 & 0.152 & 0.132 \\
        4  & 120 & 0.200 & 0.168 \\
        5  & 105 & 0.221 & 0.152 \\
        6  & 129 & 0.246 & 0.207 \\
        7  & 124 & 0.378 & 0.210 \\
        8  & 116 & 0.224 & 0.215 \\
        9  & 154 & 0.406 & 0.225 \\
        10 & 142 & 0.474 & 0.249 \\
        \hline
    \end{tabular}
\end{table}

\begin{table*}
    \centering
    \caption{Table of statistics median, median absolute deviation (MAD), and standard deviation ($\sigma$) of differences between the estimated stellar parameters (\teff, \logg, \meta) and literature values from the spectroscopic surveys LAMOST and APOGEE (validation dataset).} 
    \label{TableVDcomparation}
    \begin{tabular}{lcccccccccc}
        \hline\hline
        Neighbor-  & N & \multicolumn{3}{c}{$\Delta$ \teff\ (K)} & \multicolumn{3}{c}{$\Delta$ \logg\ (dex)} & \multicolumn{3}{c}{$\Delta$ \meta\ (dex)} \\
        hood &       & Median &  MAD & $\sigma$ & Median &  MAD & $\sigma$ & Median &  MAD & $\sigma$ \\
        \hline
1 & 6436 & 0 & 0  & 103 & 0 &  0 & 0.151 & 0 & 0 & 0.082 \\
2 & 7216 & -1 & 93  & 106 & -0.002 &  0.081 & 0.093 & -0.001 & 0.131 & 0.148 \\
3 & 56145 & -17 & 103  & 124 & -0.017 &  0.158 & 0.210 & -0.002 & 0.146 & 0.167 \\
4 & 47848 & -35 & 115  & 141 & -0.030 &  0.179 & 0.251 & -0.011 & 0.188 & 0.210 \\
5 & 20144 & -16 & 116  & 136 & -0.054 &  0.208 & 0.353 & 0.012 & 0.165 & 0.207 \\
6 & 22136 & -12 & 126  & 167 & -0.054 &  0.192 & 0.373 & 0.011 & 0.192 & 0.243 \\
7 & 45962 & 2 & 105  & 140 & -0.092 &  0.238 & 0.441 & 0.020 & 0.199 & 0.234 \\
8 & 5699 & 7 & 107  & 179 & -0.022 &  0.150 & 0.401 & 0.031 & 0.276 & 0.304 \\
9 & 4108 & -3 & 133  & 201 & -0.052 &  0.235 & 0.795 & 0 & 0.274 & 0.322 \\
10 & 3947 & 15 & 147  & 214 & -0.059 &  0.368 & 0.647 & -0.052 & 0.257 & 0.298 \\
        \hline
    \end{tabular}
\tablefoot{
Any reported statistical value (median, MAD, or sigma) with an absolute magnitude of $10^{-4}$ or lower was considered negligible and they are shown as zeros.}
\end{table*}

\begin{figure*}
    \centering
        \includegraphics[width=0.98\textwidth]{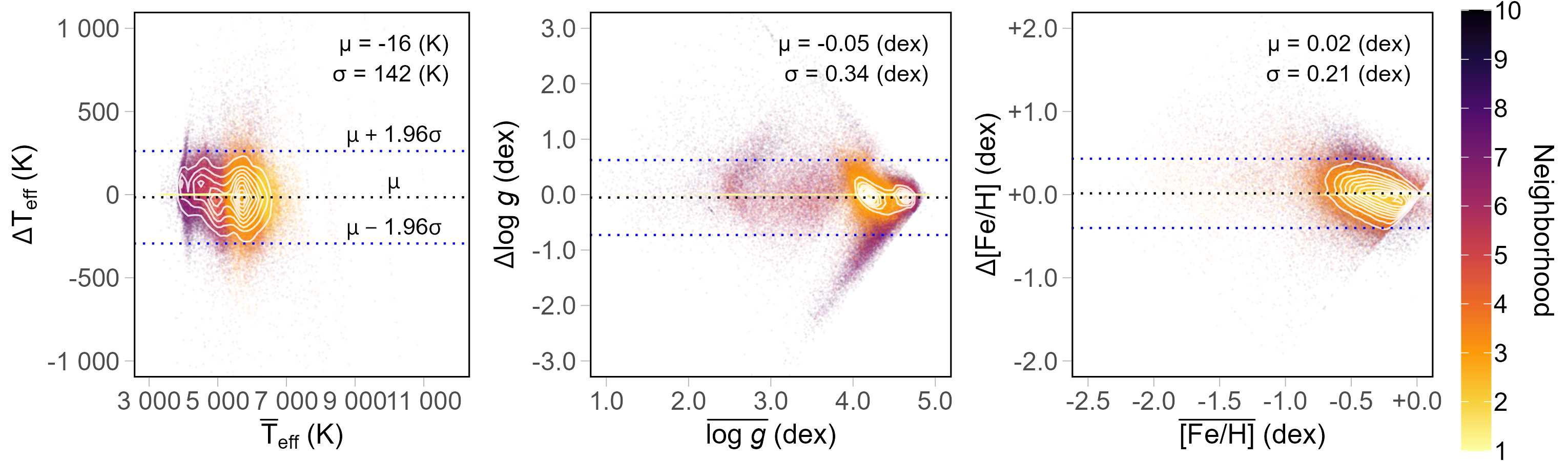}
    \caption{Bland-Altman diagrams for \teff\ ({\sl left panel}), \logg\ ({\sl middle panel}), and \meta\ ({\sl right panel}), calculated using 105 colors, in comparison to values reported for the validation data. The dotted horizontal black line represents their mean difference and the dotted horizontal blue lines are the mean difference plus or minus 1.96$\sigma$ to illustrate the confidence region. The color scheme represents the quality of the neighborhood proposed in our method, with yellow being the most reliable results (neighborhood 1) and black-purple being the less reliable ones (neighborhood 10). The white contour lines are the densities calculated over the entire sample.}
    \label{BA-Refdata}
\end{figure*}

\begin{figure*}
	\centering
    	\includegraphics[width=0.86\textwidth]
        {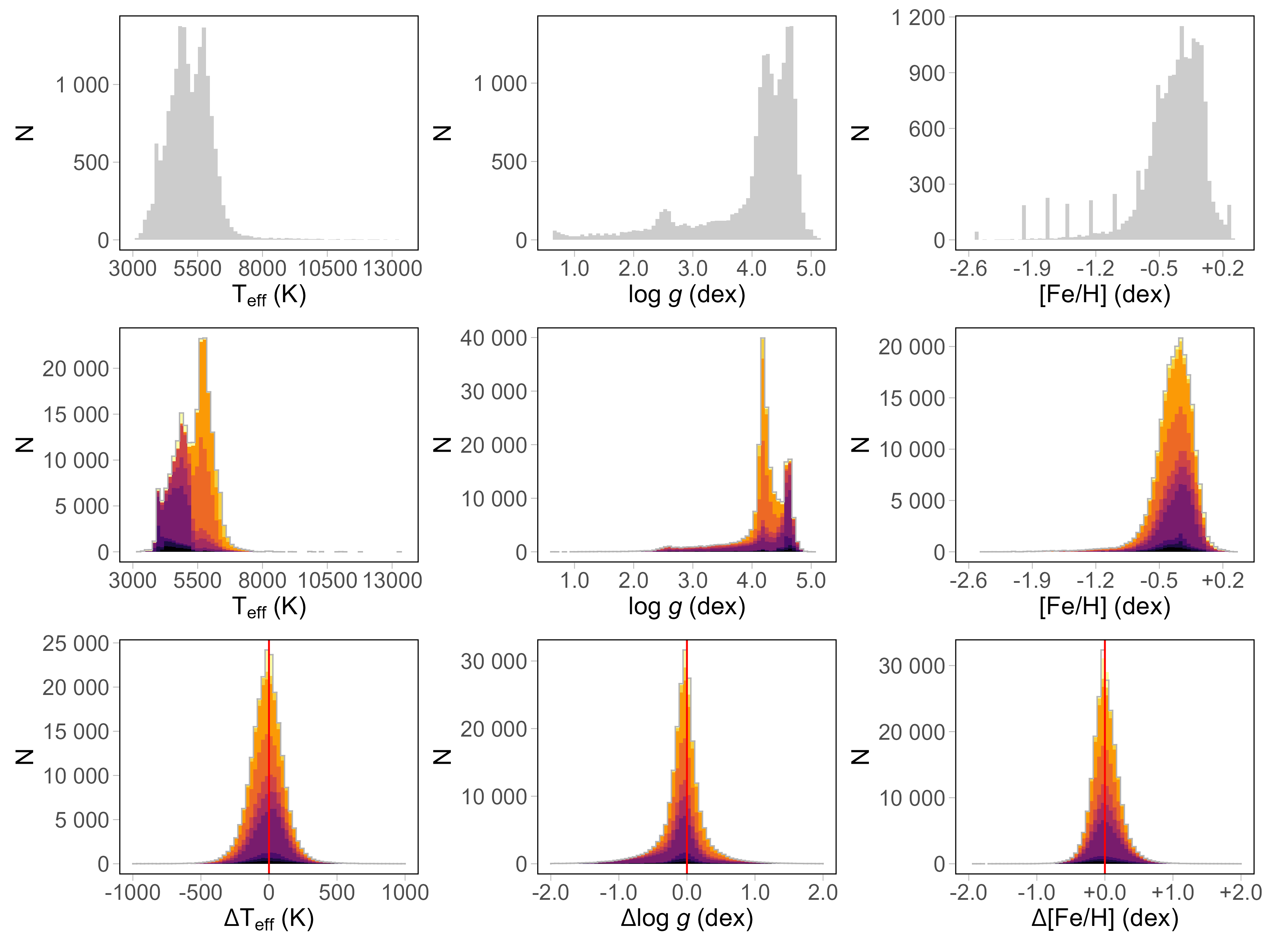}
	\caption{Distributions of the reference, estimated, and residual stellar parameters for the validation sample. The upper row shows the distributions of the stellar parameters in the reference data. The middle row shows the distributions of the parameters estimated with the 105 color method for the validation objects, separated by neighborhood value. The lower row shows the residual distributions, defined as the difference between our estimates and the spectroscopic validation values. The vertical red line marks a null residual. The color scale is the same as that shown in Fig.~\ref{BA-Refdata}.}
	\label{Histograms105vsValData}
\end{figure*}

Our method did not compute stellar parameters for 597\,965 stars. We found that none of these objects had counterparts in the reference dataset within any of the ten neighborhood radii. Without any nearby reference star for photometric comparison, it was not possible to derive their parameters. 
Notably, 98\% of them did not meet the photometric quality criteria. Their color indices fell well outside the typical range observed in stars with reliable parameter determinations. While most objects in the sample have color values between –4.56 and 12 mag, these outliers exhibited values from –20.1 to 23.7 mag. Such extreme values lack physical interpretation, reinforcing that the method effectively rejects spurious detections.

\subsection{Validation of our results}\label{SpectroscopicContrast}

To assess the reliability of our method, we compared the estimated stellar parameters (\teff, \logg, \meta) to those obtained from spectroscopy for the 219\,641 objects (LAMOST and APOGEE) in our validation sample (see details in Sect.~\ref{comparisondata}). We quantified the differences between our estimates and our validation sample by computing the median, the MAD, and the standard deviation ($\sigma$). Among them, MAD offers a robust measure of dispersion, as it is significantly less sensitive to outliers than, for example, the mean or the standard deviation. Although $\sigma$ is widely used to describe dispersion in normally distributed data, it is known to be strongly influenced by extreme values. Nevertheless, we also report $\sigma$ values due to its frequent use in the literature and because it is required for the Bland–Altman analysis described later in this section.

Table~\ref{TableVDcomparation} reports statistics of $\Delta X$\footnote{Throughout this work, $\Delta X$ = $X_{our}-X_{spec}$ for each parameter $X$ = \teff, \logg, or \meta.} by neighborhood, which shows that the median of the differences in effective temperature ($\Delta$\teff) are quite small, having 35\,K and 0\,K as the absolute maximum and minimum differences, respectively. Regarding the MAD of these differences, a value of 0\,K was obtained for the smallest neighborhood radius and reached a maximum value of 147\,K in the largest radius. $\Delta$\logg\ presented values between $-2.670 \times 10^{-8}$ and $-$0.092\,dex for the median and a MAD between $1.923 \times 10^{-7}$ and 0.368\,dex, which were assigned to neighborhoods 1 and 10, respectively. It is worth mentioning that differences of the order $10^{-4}$ or smaller were considered as zero in Table~\ref{TableVDcomparation}. $\Delta$\meta\ medians also have values close to zero, with a maximum absolute difference of 0.052\,dex. In this case the MAD values have a maximum of 0.276\,dex in neighborhood 8.

These statistical results show that smaller neighborhoods (i.e., shorter radii) tend to produce stellar parameter estimates with lower dispersion. 
This effect is more pronounced for neighborhoods 5 and below across all parameters (\teff, \logg, and \meta), with MAD values increasing with neighborhood radii. 
These findings suggest that smaller radii yield more reliable estimates and thus neighborhoods can be used as a proxy for the reliability of the derived parameters.

We used the Bland–Altman diagram to assess the agreement between our estimates and the spectroscopic measurements. This method represents the difference between two measurements as a function of their mean, allowing us to identify systematic offsets and assess the limits of agreement between two measurements \citep{haghayegh2020comprehensive}.
The limits of agreement are defined as $\mu \pm 1.96\sigma$, where $\mu$ is the mean difference (Called BIAS) and $\sigma$ its standard deviation, corresponding to a confidence interval $\sim$95\%. Figure~\ref{BA-Refdata} shows that, for the three parameters under study, the vast majority of objects lie within these limits, indicating a high level of consistency between both approaches.

For the full validation data, 96\% of the effective temperature differences fall within the confidence region, with $\sigma = 142$\,K. Similarly, 94\% and 95\% of the data for \logg\ and \meta\ are within the limits, with $\sigma = 0.34$\,dex and $0.21$\,dex, respectively. These fractions are consistent with the expected statistical behavior, confirming that our method provides reliable estimates with no significant systematic deviations.

The global values of $\sigma$ reported above should not be interpreted as a single precision level for the full catalog, since the validation sample includes objects with different photometric qualities and different neighborhood values. As shown in Table~\ref{TableVDcomparation}, the dispersion is strongly dependent on the neighborhood parameter. 

Some of these deviations, especially those observed in $\log g$, are mainly found for objects assigned to large neighborhoods, for which the local reference set may include stars with more heterogeneous physical parameters. In such cases, the degeneracy between $\log g$, $T_{\rm eff}$, and [Fe/H] is not fully removed by the photometric information alone. Therefore, estimates obtained with large neighborhood values should be interpreted with caution.

Figure~\ref{Histograms105vsValData} provides an additional diagnostic of this behavior. The upper panels show the distribution of the stellar parameters in the reference data, while the middle panels show the distribution of parameters estimated for the validation dataset, according to their neighborhood. It is worth highlighting that our estimates are contained within the reference data distribution.

The lower panels of Fig.~\ref{Histograms105vsValData} illustrate the residual distributions between our estimates and the spectroscopic validation values. The residual histograms are centered close to zero for the three stellar parameters, but their widths increase for larger neighborhoods. This confirms that the neighborhood parameter traces the reliability of the local match in the color space. 
The middle panels also show that different neighborhood values contribute to different regions of the parameter space. This is particularly relevant for $T_{\rm eff}$, where the use of several neighborhood radii allows the method to provide estimates over a broader range of temperatures. For \logg, the broader residual distribution and the structured
behavior visible in the Bland--Altman diagram indicate a stronger sensitivity to local reference-sample heterogeneity and to ambiguities in the mapping between photometric colors and atmospheric parameters, particularly for objects requiring large neighborhoods.

An important factor in this analysis is the behavior of the parameters derived for objects with HQP (HQP subset) and those that did not meet the quality criteria established in Sect.~\ref{target_data} (SCP subset), in order to validate whether this is an influential factor. The validation data contains 219\,641 stars. Of these, 211\,360 belong to the HQP subset, while the remaining 8\,281 are part of the SCP group. For HQP stars, the median differences between our estimates and the spectroscopic values were $-$14\,K for $\Delta T_{\rm eff}$, $-$0.039\,dex for $\Delta$\logg, and $-$0.001\,dex for $\Delta$[Fe/H]. The corresponding MAD values were 73\,K, 0.117\,dex, and 0.115\,dex, respectively.

The same statistical comparison was performed for stars in the SCP subset. The results showed a slight increase in dispersion: the median differences were 61\,K for $\Delta T_{\rm eff}$, 0.047\,dex for $\Delta \log{g}$, and $-$0.096\,dex for $\Delta$\meta, with corresponding MAD values of 107\,K, 0.170\,dex, and 0.136\,dex, respectively.

Among SCP stars located in neighborhoods with values five or lower, the results exhibit even lower dispersion. The MAD values in this case were 106\,K for $\Delta T_{\rm eff}$, 0.120\,dex for $\Delta \log{g}$, and 0.109\,dex for $\Delta$\meta, which are similar to those reported in Table~\ref{TableVDcomparation}. These values indicate that our method is still capable of recovering reliable stellar parameters for objects that do not satisfy all photometric quality criteria, with a dispersion comparable to that obtained for the HQP subset.

One of the objectives of this work was to assess the reliability of the inferred stellar parameters for objects with somewhat compromised photometry. We have a low percentage ($\sim$4\%) of this type of data in the validation dataset; however, 66\% of the sample with derived stellar parameters is part of the SCP subset. Therefore, in the next section we present a further discussion concerning the parameter uncertainties for objects from the SCP subset compared to the results obtained by other works.

\section{Discussion}\label{discuss}

\subsection{Comparison with previous results}\label{Other authors}

We compared our results with previous studies that analyzed J-PLUS data to estimate stellar atmospheric parameters such as \teff, \logg, and \meta\ using photometry. This comparison focuses on the works of \citet{Wang2022} and \citet{Yang2024}, as introduced in Sect.~\ref{comparisondata}.

We adopted the same approach to compare the estimates by computing the differences between our estimates and those reported in these studies, using the full set of 105 colors. The comparison with \citet{Wang2022} yields a median difference of 6\,K in \teff and MAD of 76\,K. The comparison with \citet{Yang2024} results in a median \teff\ difference of 35\,K and a MAD of 66\,K.

Concerning \logg, the MAD values are 0.096\,dex and 0.144\,dex for comparisons with \citet{Wang2022} and \citet{Yang2024}, respectively. For \meta, the MAD values are 0.128\,dex and 0.174\,dex.

\begin{figure*}
    \centering
        \label{HistoDIFF105vsAUTHORSall}
            \centering
    \includegraphics[width=0.95\textwidth]{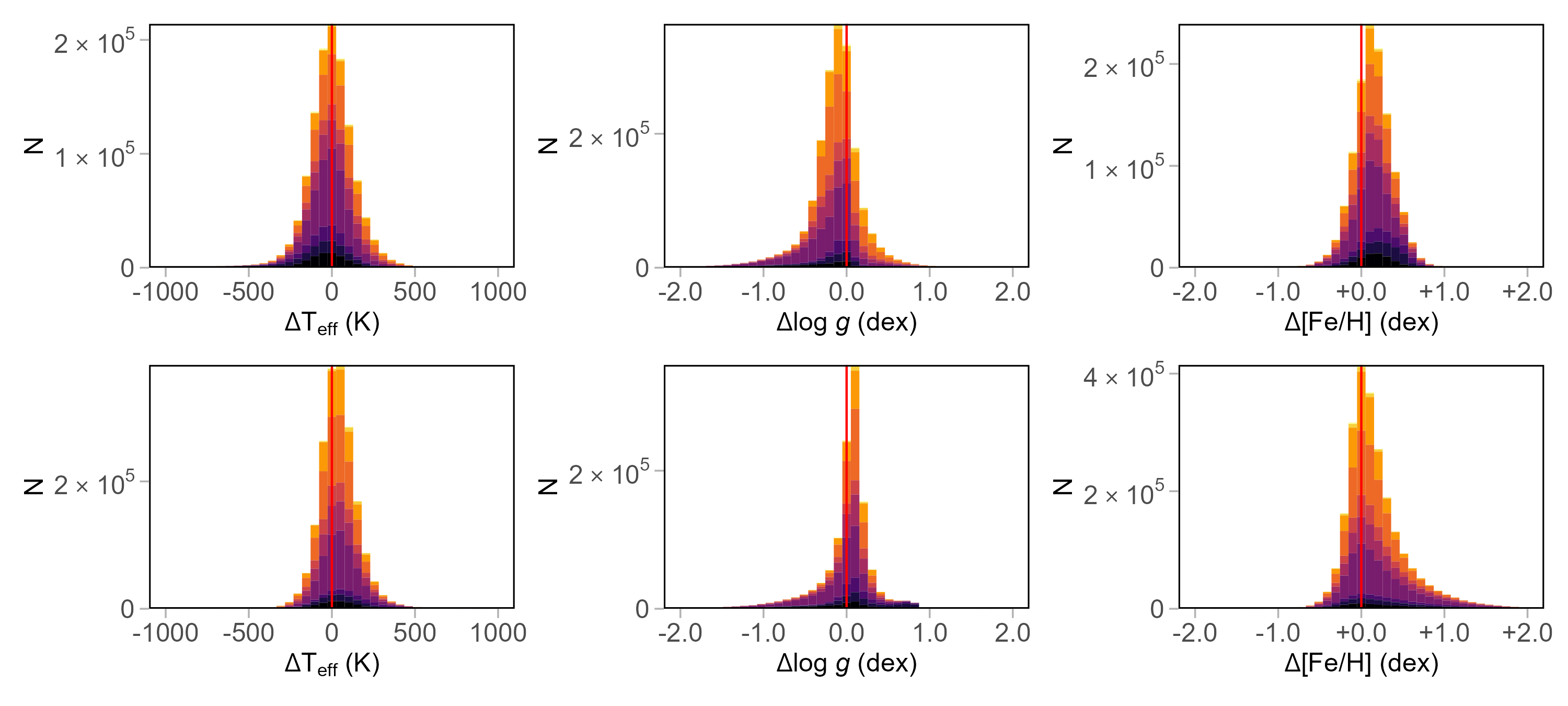}    
            \caption{Histograms of the differences in \teff ({\sl left panels}), \logg ({\sl middle panels}), and \meta\ ({\sl right panels}) between our results, obtained with 105 colors, and those from \citet{Wang2022} ({\sl top panels}) and \citet{Yang2024} ({\sl bottom panels}). Vertical red lines represent a null difference. The color bar is the same as the one shown in Fig.~\ref{BA-Refdata}.}
            \label{HistoDIFF105vsOTHERSworks}
\end{figure*}

Table~\ref{statsOthers} shows the statistics of all differences, separated by neighborhood and by photometric quality (HQP and SCP; see Sect.~\ref{target_data}). In general, the differences found in \teff\ for both samples, \citet{Wang2022} and \citet{Yang2024}, are similar, which shows the capability of our method of deriving stellar \teff\ regardless of possible limitations in the J-PLUS photometry. Although higher differences are found for \logg\ and \meta\ derived for the SCP sample with larger neighborhoods, the estimated values show less dispersion for smaller neighborhoods.

Figure~\ref{HistoDIFF105vsOTHERSworks} illustrates the differences mentioned above for the estimated parameters, color-coded by neighborhood, where darker colors represent larger neighborhoods. The highest dispersions are associated with objects in the largest neighborhoods. Based on the criterion $\mu \pm 1.96\sigma$  (see Sect.~\ref{SpectroscopicContrast}), 96\% of the sample falls within the confidence interval for $\Delta$\teff\ in both comparison sets. For $\Delta$\logg, 94\% of the stars are within the expected range, and for $\Delta$\meta, the coverage is 95\% in both cases. 
These results indicate a good level of agreement between our estimated stellar parameters and those derived by \citet{Wang2022} and \citet{Yang2024}.

Despite the potential of photometric information to infer stellar parameters, previous studies based on J-PLUS have also reported limitations in their determination.\citet{Wang2022} found that surface gravity exhibited the worst normalized performance among the three inferred parameters, with a blind-test RMSE of 0.35\,dex, increasing to 0.65\,dex for sources in the extrapolation regime. Furthermore, in comparison with \citet{Yang2022}, \citet{Wang2022} identified extreme and physically implausible values, reaching 882.44 and -237.24\,dex for \logg, and 460.39 and -567.16\,dex for \meta, suggesting that this may be associated with objects located near the feature space boundaries and overfitting in the artificial neural network. On the other hand, \citet{Yang2024} reported good agreement between their surface gravity estimates and those of LAMOST, with a shift of -0.02\,dex and a dispersion of 0.11\,dex. However, they found that photometric metallicities become unreliable for targets whose colors have low similarity to those represented in the training set, for flg $[Fe/H] < 0.60$\,dex, the dispersion exceeds 0.8\,dex. Taken together, these results indicate that good overall agreement does not preclude errors for individual sources located in poorly represented regions of the parameter space, further supporting the use of our neighborhood index as a source-by-source indicator of the reliability of derived parameters.

Figure~\ref{CostrastALLAUTHORS} (Appendix~\ref{AllComparations}) compares the stellar atmospheric parameters estimated using our method based on 105 colors for 1\,236 stars common to all datasets considered in this work: LAMOST, APOGEE, \citet{Wang2022}, and \citet{Yang2024}. This figure shows that the comparison between our estimates and the spectroscopic catalogs does not yield similar distributions for LAMOST and APOGEE. This behavior is expected, since the two surveys rely on different spectral ranges, spectral resolutions, observing strategies, and stellar-parameter pipelines. Therefore, systematic differences between their atmospheric-parameter scales may be present. 
In our approach, the impact of such differences is mitigated by estimating the parameters locally in the color space and quantifying their uncertainty with the estimated error for each parameter.

We compared the common objects between LAMOST and APOGEE using the same parameter intervals and quality criteria adopted for the reference sample. We found median absolute deviations of  $MAD({T_{\rm eff,LAMOST}-T_{\rm eff,APOGEE}})$ = 90\,K, $MAD({\log{g}_{LAMOST}-\log{g}_{APOGEE}})$ = 0.112 dex, and $MAD({[Fe/H]_{LAMOST}-[Fe/H]_{APOGEE}})$ = 0.058 dex. These values are of the same order as the MAD values reported in Tab.~\ref{TableVDcomparation} and the RMS error (RMSE) values reported in Fig.~\ref{CostrastALLAUTHORS}, suggesting that the differences between the LAMOST and APOGEE scales are comparable to the internal dispersion obtained in our validation analysis.

Linear regression analysis was used as a diagnostic tool to evaluate the agreement between our results and the reference datasets. The fits were computed using weighted least-squares (WLS), with the inverse squared uncertainties of the reference parameters as weights. 

As previously observed, larger neighborhood values tend to result in greater dispersion in parameter estimates. In contrast, stars associated with neighborhood indices of one or two show tightly clustered distributions, suggesting more consistent and stable predictions in those regions.
Reliable results are expected to yield correlation coefficients ($R^2$) near unity and intercepts close to zero. Comparisons with spectroscopic measurements (first and second rows) show regression slopes near one, confirming the consistency of our method with external reference data. 
The narrow confidence intervals of the fitted relations reflect the strong clustering of the data around the best-fit lines and the concentration of most stars within well-populated regions of parameter space\footnote{This subset is concentrated in $4500 \leqslant T_{\rm eff} \leqslant 5000$, $4 \leqslant \log{g} \leqslant 5$, and $-0.5 \leqslant [Fe/H] \leqslant 0$.}. The low RMSE values, reported in each panel, confirm the robustness of the agreement in the ranges of explored parameters. 
Comparisons with the photometric estimates of \citet{Wang2022} and \citet{Yang2024} (third and fourth rows) also exhibit high $R^2$ values for \teff and \logg, with moderate correlations for \meta.

Our method estimates stellar parameters that agree with literature values of \citet[][]{Wang2022} and \citet[][]{Yang2024}, and shows results of the same order, denoting low uncertainties. In contrast to those studies, our approach requires no previous training, greatly reducing computation time. 

An important advantage of our method is that it inherently limits the estimation of stellar parameters to regions of the photometric feature space that are supported by reference data. In contrast to supervised approaches (for example, neural networks), which may extrapolate in sparsely populated regions, our neighborhood-based strategy only assigns parameters when reference stars with known parameters are found within a defined local radius. This constraint reduces the possibility of spurious solutions and provides a quality indicator through the neighborhood parameter.

\subsection{Color sample reduction}\label{Color_sample_reduction}

Reducing the number of photometric colors while maintaining the ability to estimate atmospheric parameters effectively would significantly speed up computations and enable the analysis of a larger sample -- photometric surveys tend to have stars that do not have magnitudes available in all photometric bands. Many of the 105 colors used in our full model exhibit a similar behavior with respect to stellar parameters and can be grouped accordingly. Appendix~\ref{MatrixCoor} (Fig.~\ref{Coor}) shows the color correlation matrix, which illustrates this redundancy. We then applied principal component analysis (PCA), an algorithm that identifies correlated variables and reduces dimensionality.

Considering the eigenvalues, which represent the variance explained by each principal component, we defined a smaller subset of colors based on those that had similar loadings on the most significant components. The fact that colors with similar loadings exhibit similar behavior can be validated by a high correlation index between them.

The explained variance takes the largest values in the first four components with an explained cumulative variance of 96\%. Considering these four components,  the colors formed 11 groups of similar behavior, obtained by applying the Hartigan test to the estimated eigenvalues for each color in the four PCA components. 

We selected the color with the highest correlation index with respect to the other colors within each of the 11 groups. The 11 selected colors are $J0378-J0660$, $J0395-g$, $J0410-J$, $J0430-i$, $J0430-z$, $g-K$, $J0660-H$, $i-z$, $i-J$, $J0861-z$, and $z-H$.
The use of this reduced set expanded the number of analyzable stars to 6,539,507 (from the original pool of 7\,400\,214 after applying \texttt{SExtractor}'s \texttt{CLASS\_STAR} criterion; see Sect.~\ref{target_data}). Stellar parameters were successfully derived for 6\,045\,593 stars, an increase of 397\,107 objects compared to the 105 color set. Additionally, computation time decreased by more than 50\%.

\begin{figure}
	\centering
	\includegraphics[width=0.93\hsize]{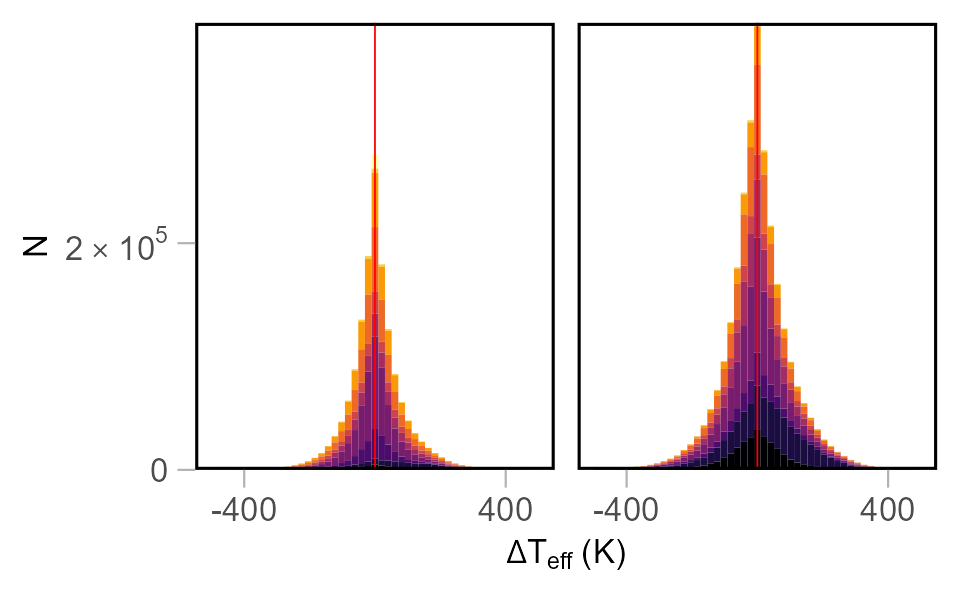}
    \includegraphics[width=0.93\hsize]{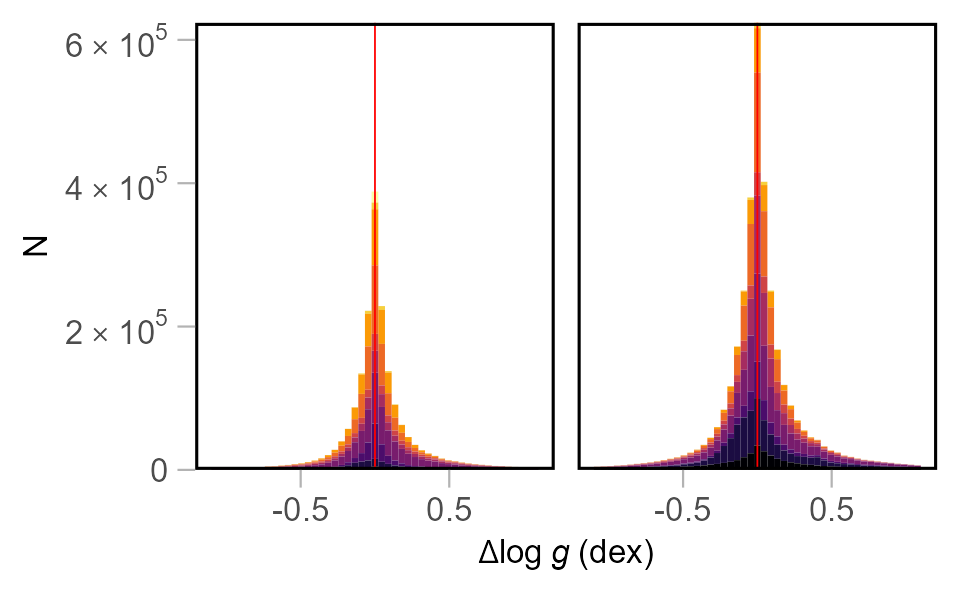}
    \includegraphics[width=0.93\hsize]{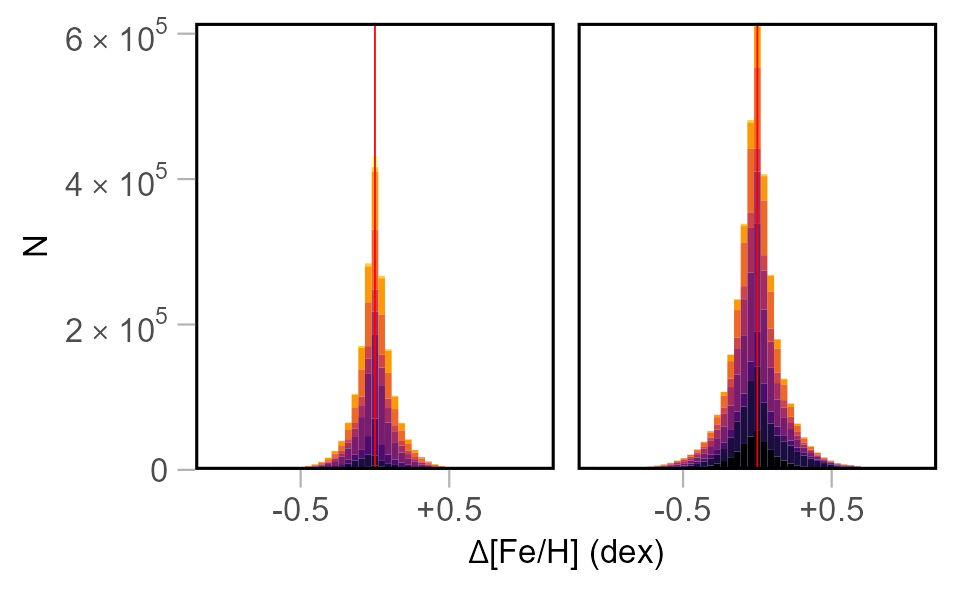}
	\caption{Histograms of the differences in \teff, \logg, and \meta, from top to bottom, respectively, obtained with our method using 105 colors and the results when using this same method with the 11 chosen colors. Left panels show the differences obtained for the HQP subset and right panels for those in the SCP subset. The vertical red line marks a null residual. The color bar is the same as the one shown in Fig.~\ref{BA-Refdata}.}
	\label{HistoDIFF105vs11}
\end{figure}

Figure~\ref{HistoDIFF105vs11} shows the distribution of the parameter differences between the results obtained with 105 and the reduced set of 11 colors. The left panels show the objects in the HQP subset, which present smaller differences between the parameters, compared to the SCP subset (right panels). The smallest discrepancies are observed for HQP stars in smaller neighborhoods. 
The median differences for the HQP subset are null for \teff\ and \meta, and are equal to 0.002\,dex for \logg.
The SCP subset also shows minimal median differences in \teff, with slightly larger deviations in \logg and \meta, 
although all differences remain close to zero. Our results indicate that even after applying a significant dimensionality reduction and despite limitations in the photometric quality, we were able to recover similar derived stellar parameters.

For HQP stars, the MAD values are 39\,K, 0.084\,dex, and 0.064\,dex for \teff, \logg, and \meta, respectively. For SCP stars, these increase slightly to 50\,K, 0.099\,dex, and 0.086\,dex. It should be emphasized that these values were estimated considering all neighborhoods and that the differences between both samples are better noted when comparing individual neighborhoods.

Similar to trends in the validation dataset, greater dispersion is seen in larger neighborhoods, especially in the SCP subset. For instance, the best-case scenario (HQP stars with neighborhood 1) yields MAD values of 0 K, 0 dex, and 0 dex for \teff, \logg, and \meta, respectively. In contrast, the worst-case scenario (SCP stars with neighborhood 10) shows MAD values of 51 K, 0.211 dex, and 0.092 dex.

In summary, the applied color reduction has proven to be a good reference for stellar parameters, allowing a larger dataset to be analyzed, despite some increase in the derived uncertainties when compared to the results obtained with 105 colors and the validation data. This implementation generated a significant reduction in the computational time required compared to the time invested using 105 colors. 

\section{Conclusions}\label{concl}

Spectroscopy provides more direct parameter constraints; however, the extensive availability of photometric surveys offers a valuable alternative when spectroscopic data are not available.
We estimated the atmospheric stellar parameters (\teff, \logg, \meta) for a total of 5\,648\,486 stars in J-PLUS DR3, using 105 colors. In this work, we integrated photometric data from 12 J-PLUS bands and three 2MASS magnitudes. Unlike previous studies with similar objectives using J-PLUS data, we incorporated near-infrared data to enhance parameter inference.

The adopted method was based on the k-means method combined with a modified KNN algorithm. 
Our approach consisted of a comparative analysis between reference datasets comprising both observational data with spectroscopically derived atmospheric stellar parameters and synthetic data from models. As reference, we adopted data from LAMOST, APOGEE, and MIST models to infer \teff, \logg, and \meta\ for the objects in our sample. Our method is computationally efficient, covers a broad range of stellar parameters, scales well to large datasets, and avoids the need for a conventional, training-intensive supervised learning stage.

Considering all 105 colors, our results were in agreement with the spectroscopic estimates from LAMOST and APOGEE. The estimated parameters were obtained not only for data with HQP but also for objects with somewhat compromised photometry. This classification allowed for a systematic assessment of the robustness of our method under varying observational conditions. Our results demonstrate that the method yields reliable atmospheric parameters even when the photometric data are somewhat compromised.

The analysis showed that adopting different neighborhood radii enabled the determination of stellar parameters not only for objects with nearby reference data for which a small radius (neighborhood 1) was sufficient but also for cases requiring progressively larger radii (up to neighborhood 10). A comparison with a validation dataset, as well as with results reported in the literature, indicated that the neighborhood size correlates with the reliability of our estimates. In particular, the most reliable parameters are generally obtained for neighborhoods with radii less than or equal to five.

The neighborhood index and the reported parameter error provide complementary information. The former describes the extent of the color space that had to be searched to identify reference support, whereas the latter quantifies the dispersion among the stellar-parameter labels of the reference objects actually used for each estimate.

We reviewed the efficacy of using numerous color indices for parameter estimation while exploring the possibility of identifying a smaller set of colors that could produce parameter estimates comparable in accuracy to those obtained from the complete set. 
Dimensionality reduction via PCA revealed that 11 selected colors retained the essential information required for parameter estimation. This reduced set of colors enabled the analysis of a larger number of stars while reducing computational time. Despite minor increases in estimated uncertainties, the reduced model maintains consistency with results from the full 105-color analysis.

\section{Data availability}\label{CDS}

The full version of Table~\ref{105Results} is available at the CDS via \url{http://cdsweb.u-strasbg.fr/cgi-bin/qcat?J/A+A/11111111111}.

\begin{acknowledgements}
J.A.'s contribution to this work is part of his academic activities as a professor at Universidad Militar Nueva Granada, Bogotá, Colombia. 
J.A., P.Cruz, E.S. and FJE acknowledge support from the Spanish Virtual Observatory (\url{https://svo.cab.inta-csic.es}) project funded by the Spanish Ministry of Science, Innovation and Universities /State Agency of Research MCIU/AEI/10.13039/501100011033 through grant PID2023-146210NB-I00. 
P.Cruz also acknowledges support from grant PID2019-107061GB-C61.  
P.Coelho acknowledgments the support by Conselho Nacional de Desenvolvimento Científico e Tecnológico (Process 308715/2025-0). 
A.E. acknowledges the financial support from the Spanish Ministry of Science and Innovation and the European Union - NextGenerationEU through the Recovery and Resilience Facility project ICTS-MRR-2021-03-CEFCA.
The work of V.M.P. is supported by NOIRLab, which is managed by the Association of Universities for Research in Astronomy (AURA) under a cooperative agreement with the U.S. National Science Foundation.
A.A.C. acknowledges financial support from the Severo Ochoa grant CEX2021-001131-S funded by MCIN/AEI/10.13039/501100011033 and the project PID2023-153123NB-I00 funded by MCIN/AEI.

Based on observations made with the JAST80 telescope and T80Cam camera for the J-PLUS project at the Observatorio Astrof\'{\i}sico de Javalambre (OAJ), in Teruel, owned, managed, and operated by the Centro de Estudios de F\'{\i}sica del Cosmos de Arag\'on (CEFCA). We acknowledge the OAJ Data Processing and Archiving Unit (UPAD;
\citealt{upad}) for reducing the OAJ data used in this work. Funding for the J-PLUS Project has been provided by the Governments of Spain and Arag\'on through the Fondo de Inversiones de Teruel; the Aragonese Government through the Research Groups E96, E103, E16\_17R, E16\_20R, and E16\_23R; the Spanish Ministry of Science and Innovation (MCIN/AEI/10.13039/501100011033 y FEDER, Una manera de hacer Europa) with grants PID2021-124918NB-C41, PID2021-124918NB-C42, PID2021-124918NA-C43, and PID2021-124918NB-C44; the Spanish Ministry of Science, Innovation and Universities (MCIU/AEI/FEDER, UE) with grants PGC2018-097585-B-C21 and PGC2018-097585-B-C22; the Spanish Ministry of Economy and Competitiveness (MINECO) under AYA2015-66211-C2-1-P, AYA2015-66211-C2-2, AYA2012-30789, and ICTS-2009-14; and European FEDER funding (FCDD10-4E-867, FCDD13-4E-2685). The Brazilian agencies FINEP, FAPESP, and the National Observatory of Brazil have also contributed to this Project.
This publication makes use of data products from the Two Micron All Sky Survey, which is a joint project of the University of Massachusetts and the Infrared Processing and Analysis Center/California Institute of Technology, funded by the National Aeronautics and Space Administration and the National Science Foundation. 
This research has made use of TOPCAT \citep{Taylor05}.

\end{acknowledgements}

   \bibliographystyle{aa} 
   \bibliography{StellarParameters1.bib}

\begin{appendix}
\nolinenumbers
\section{Additional material}\label{AppenA}

Table \ref{Kmeansphase1} shows the distribution of the reference data in each of the 19 clusters suggested by the Hartigan test implementation. The first and second columns shows the cluster number and the source, respectively. The third column indicates the number of objects (observed or synthetic) within each cluster. The remaining columns shows the minimum, the maximum, and the median values of \teff, \logg, and \meta, respectively, as the summary of the distribution. A detailed information is given in Sect.~\ref{MLmethodology}.

\begin{table*}
    \caption{Distribution of the reference sample into 19 clusters.} 
    \label{Kmeansphase1}
    \centering
    \begin{tabular}{lcccccccccccccc}
        \hline
        \hline
        Cluster & Source & N & \multicolumn{3}{c}{\teff\ (K)} & \multicolumn{3}{c}{\logg\ (dex)} & \multicolumn{3}{c}{\meta\ (dex)} \\
        \cline{4-6} \cline{7-9} \cline{10-12}
        & & & Min & Max & Median & Min & Max & Median & Min & Max & Median \\
        \hline
1 & LAMOST & 4 & 5742 & 11720 & 8339 & 4.029 & 4.388 & 4.123 & $-$0.44 & $-$0.026 & $-$0.174 \\
1 & MIST & 290 & 7968 & 18766 & 10277 & 1.099 & 4.554 & 3.461 & $-$4 & 0.5 & $-$2 \\
2 & APOGEE & 880 & 3186 & 4645 & 3815 & 0.652 & 5.067 & 4.655 & $-$1.435 & 0.303 & $-$0.054 \\
 & LAMOST & 572 & 3744 & 5988 & 3952 & 1.084 & 4.897 & 4.596 & $-$1.387 & $-$0.001 & $-$0.388 \\
 & MIST & 2568 & 2212 & 4661 & 3553 & -1.093 & 5.416 & 0.243 & $-$4 & 0.5 & $-$0.75 \\
3 & MIST & 3 & 167112 & 168845 & 167740 & 6.812 & 6.972 & 6.882 & $-$3 & $-$1.75 & $-$2 \\
4 & MIST & 38 & 99452 & 108156 & 103784 & 5.501 & 7.355 & 6.375 & $-$4 & 0.5 & $-$0.625 \\
5 & MIST & 52 & 89612 & 98827 & 93157 & 5.201 & 7.608 & 6.969 & $-$4 & 0.5 & $-$0.875 \\
6 & MIST & 22 & 128097 & 138551 & 131921 & 5.958 & 7.390 & 6.741 & $-$3 & 0 & $-$1.375 \\
7 & APOGEE & 953 & 4560 & 6305 & 5517 & 2.070 & 4.595 & 4.200 & $-$1.589 & 0.295 & $-$0.272 \\
 & LAMOST & 5457 & 4888 & 6889 & 5760 & 1.805 & 4.890 & 4.250 & $-$2.441 & 0.03 & $-$0.33 \\
 & MIST & 522 & 4629 & 6034 & 5173 & 0.126 & 4.957 & 2.392 & $-$4 & 0.5 & $-$1.5 \\
8 & MIST & 10 & 151161 & 161093 & 156536 & 6.325 & 6.985 & 6.657 & $-$3 & 0.5 & $-$1.375 \\
9 & MIST & 162 & 17003 & 33645 & 26547 & 2.351 & 7.972 & 5.614 & $-$4 & 0.5 & $-$1.25 \\
10 & MIST & 10 & 140204 & 149902 & 145439 & 6.441 & 7.295 & 6.746 & $-$3 & 0.5 & $-$1 \\
11 & APOGEE & 2932 & 3956 & 5558 & 4822 & 0.952 & 4.753 & 3.975 & $-$1.945 & 0.324 & $-$0.188 \\
 & LAMOST & 3998 & 3852 & 6355 & 4804 & 0.932 & 4.900 & 4.618 & $-$2.41 & 0.059 & $-$0.288 \\
 & MIST & 676 & 3933 & 5464 & 4547 & 0.059 & 5.419 & 2.063 & $-$4 & 0.5 & $-$1.25 \\
12 & MIST & 38 & 118267 & 127575 & 123348 & 5.672 & 7.340 & 6.600 & $-$4 & 0.5 & $-$0.75 \\
13 & MIST & 88 & 38183 & 59578 & 52855 & 3.818 & 7.899 & 7.534 & $-$4 & 0.5 & $-$0.875 \\
14 & MIST & 34 & 109044 & 117759 & 113214 & 5.649 & 7.389 & 6.942 & $-$4 & 0.25 & $-$1.25 \\
15 & MIST & 135 & 29174 & 48434 & 38556 & 3.239 & 7.925 & 4.128 & $-$4 & 0.5 & $-$1.25 \\
16 & MIST & 63 & 46983 & 79817 & 74664 & 3.980 & 7.663 & 5.220 & $-$4 & 0.5 & $-$0.75 \\
17 & APOGEE & 3 & 4843 & 6002 & 5765 & 3.700 & 4.241 & 4.099 & $-$0.807 & $-$0.098 & $-$0.613 \\
 & LAMOST & 756 & 4813 & 9820 & 6441 & 3.359 & 4.468 & 4.158 & $-$2.301 & 0.001 & $-$0.363 \\
 & MIST & 654 & 5402 & 8622 & 6501 & 0.364 & 4.774 & 3.802 & $-$4 & 0.5 & $-$1.75 \\
18 & MIST & 66 & 80363 & 89401 & 84705 & 5.048 & 7.708 & 6.972 & $-$4 & 0.5 & $-$1.25 \\
19 & MIST & 79 & 46637 & 69723 & 64639 & 4.026 & 7.791 & 4.852 & $-$4 & 0.5 & $-$0.5 \\
        \hline
    \end{tabular}
    \tablefoot{The summary of their stellar parameters is given by the minimum (Min), the maximum (Max), and the median values within each group.} 
\end{table*}

Table~\ref{105Results} summarizes the estimated stellar atmospheric parameters (\teff,\logg, \meta) calculated using our method with 105 and 11 colors. The first column lists the J-PLUS names of each object. The second and third columns show their corresponding J2000 equatorial coordinates. Columns from four to nine show the stellar parameters estimated using our method with 105 colors and their corresponding errors. Columns 12 to 17 report the same parameters estimated using 11 colors and their errors. Columns 10 and 18 report the neighborhood values assigned by our method, and columns 11 and 19 indicate the number of stars from the reference dataset used to calculate the stellar parameters, for 105 and 11 colors ($N_{\rm ref}$), respectively. Column 20 reports the photometric quality subset of the object (see Sect. \ref{target_data} for details). The complete version of this table, with the derived parameters for the almost five million objects, is available online in VizieR\footnote{\url{https://vizier.cds.unistra.fr}}.

    \begin{sidewaystable*}
    \caption{Stellar parameters estimated using the complete and the reduced sets with 105 and 11 colors, respectively.}
    \label{105Results}
    \centering
   \resizebox{\textwidth}{!}{
\begin{tabular}{cccccccccccccccccccc}
  \hline\hline
  J-PLUS--ID & RA & Dec & \teff $_{,105}$ & $\sigma_{T_{\rm eff,105}}$ & \logg $_{105}$ &  $\sigma_{\log g_{105}}$ &  \meta $_{105}$ & $\sigma_{{\rm [Fe/H]}_{105}}$ & Neighbor- & $N_{{\rm ref},105}$ & \teff $_{,11}$ & $\sigma_{T_{\rm eff,11}}$ & \logg $_{11}$ &  $\sigma_{\log g_{11}}$ &  \meta $_{11}$ & $\sigma_{{\rm [Fe/H]}_{11}}$ & Neighbor- & $N_{{\rm ref},11}$ & Quality \\
    & (deg) & (deg) & (K) & (K) & (dex) & (dex) & (dex) & (dex) & hood$_{105}$ & & (K) & (K) & (dex) & (dex) & (dex) & (dex) & hood$_{11}$ &  & sample \\
  \hline
  100003-1000 & 38.37844 & 6.79366 & 3930 & 208 & 4.489 & 0.635 & $-$0.239 & 0.281 & 9 & 1217 & 3900 & 109 & 4.632 & 0.307 & $-$0.029 & 0.207 & 5 & 1 & SCP \\
  100003-10085 & 37.89399 & 7.35927 & 5310 & 140 & 4.384 & 0.304 & $-$0.307 & 0.181 & 4 & 76 & 5224 & 101 & 4.069 & 0.168 & $-$0.295 & 0.132 & 2 & 1 & HQP \\
  100003-10091 & 37.83760 & 7.36556 & 3586 & 215 & 4.662 & 0.274 & $-$0.039 & 0.095 & 9 & 245 & 3456 & 113 & 4.721 & 0.252 & $-$0.036 & 0.073 & 5 & 23 & SCP \\
  100003-10107 & 38.25897 & 7.35027 & 4597 & 109 & 4.549 & 0.273 & $-$0.152 & 0.163 & 7 & 200 & 4592 & 106 & 4.299 & 0.701 & $-$0.110 & 0.151 & 9 & 108 & HQP \\
  100003-10144 & 38.98693 & 7.36944 & 3477 & 115 & 4.716 & 0.234 & $-$0.010 & 0.081 & 9 & 108 & 3532 & 350 & 4.502 & 0.359 & $-$0.048 & 0.101 & 8 & 3 & SCP \\
  100003-10149 & 38.16998 & 7.36824 & 5623 & 132 & 4.276 & 0.163 & $-$0.400 & 0.189 & 4 & 50 & 5649 & 92 & 4.256 & 0.095 & $-$0.378 & 0.191 & 3 & 7 & SCP \\
  100003-10182 & 37.82243 & 7.37248 & 3983 & 181 & 4.515 & 0.544 & $-$0.267 & 0.282 & 9 & 1043 & 3832 & 97 & 1.625 & 2.114 & $-$1.325 & 0.717 & 6 & 220 & SCP \\
  100003-10187 & 39.12102 & 7.36652 & 5989 & 178 & 4.167 & 0.152 & $-$0.457 & 0.210 & 3 & 32 & 6055 & 168 & 4.198 & 0.226 & $-$0.373 & 0.196 & 2 & 12 & HQP \\
  100003-10189 & 37.99008 & 7.35945 & 5584 & 77 & 4.332 & 0.228 & $-$0.032 & 0.144 & 3 & 10 & 5573 & 78 & 4.315 & 0.240 & $-$0.072 & 0.150 & 2 & 8 & HQP \\
  100003-10203 & 38.62167 & 7.37408 & 3497 & 133 & 4.691 & 0.302 & $-$0.013 & 0.073 & 9 & 181 & 3415 & 177 & 4.347 & 0.858 & $-$0.018 & 0.171 & 5 & 2 & SCP \\
  100003-10223 & 38.79858 & 7.37208 & 4942 & 123 & 4.609 & 0.055 & $-$0.101 & 0.141 & 7 & 2 & 4855 & 122 & 4.648 & 0.530 & $-$0.201 & 0.217 & 9 & 1 & SCP \\
  100003-10229 & 38.94176 & 7.37336 & 4395 & 108 & 4.598 & 0.276 & $-$0.141 & 0.207 & 7 & 79 & 4400 & 118 & 4.618 & 0.052 & $-$0.061 & 0.107 & 7 & 3 & HQP \\
  100003-10245 & 38.13169 & 7.37494 & 3864 & 153 & 4.727 & 0.409 & $-$0.896 & 0.228 & 9 & 1 & 3546 & 122 & 4.851 & 0.530 & $-$0.090 & 0.217 & 9 & 1 & SCP \\
  100003-10288 & 38.20622 & 7.37929 & 4964 & 118 & 3.280 & 0.729 & $-$0.513 & 0.208 & 7 & 125 & 4978 & 119 & 3.412 & 0.723 & $-$0.474 & 0.172 & 9 & 229 & HQP \\
  100003-10294 & 38.42641 & 7.37885 & 4245 & 113 & 4.698 & 0.255 & $-$0.200 & 0.164 & 6 & 1 & 4302 & 126 & 4.649 & 0.098 & $-$0.279 & 0.164 & 3 & 26 & HQP \\
  100003-10298 & 38.26707 & 7.37838 & 4260 & 120 & 4.617 & 0.098 & $-$0.117 & 0.152 & 7 & 69 & 4242 & 36 & 4.642 & 0.071 & $-$0.049 & 0.077 & 3 & 3 & HQP \\
  100003-10304 & 38.06631 & 7.37158 & 5758 & 47 & 4.173 & 0.099 & $-$0.339 & 0.132 & 3 & 9 & 5750 & 119 & 4.249 & 0.247 & $-$0.398 & 0.125 & 2 & 7 & HQP \\
  100003-10309 & 37.93113 & 7.37490 & 3714 & 145 & 4.616 & 0.306 & $-$0.064 & 0.089 & 8 & 129 & 3699 & 135 & 4.638 & 0.256 & $-$0.064 & 0.091 & 4 & 96 & HQP \\
  100003-10332 & 38.77928 & 7.35694 & 6130 & 80 & 4.238 & 0.075 & $-$0.484 & 0.090 & 6 & 5 & 6227 & 180 & 4.179 & 0.115 & $-$0.346 & 0.184 & 4 & 151 & SCP \\
  100003-10336 & 38.33761 & 7.38360 & 3733 & 121 & 4.130 & 0.235 & $-$0.128 & 0.236 & 8 & 1 & 3733 & 143 & 4.130 & 0.267 & $-$0.128 & 0.249 & 4 & 1 & SCP \\
  100003-10348 & 38.27514 & 7.38509 & 3641 & 162 & 4.473 & 0.297 & $-$0.098 & 0.126 & 9 & 7 & 3501 & 46 & 4.424 & 0.105 & $-$0.051 & 0.098 & 5 & 2 & SCP \\
  100003-10352 & 38.05742 & 7.38375 & 3881 & 121 & 4.577 & 0.235 & $-$0.415 & 0.236 & 8 & 1 & 3881 & 143 & 4.577 & 0.267 & $-$0.415 & 0.249 & 4 & 1 & SCP \\
  100003-10360 & 39.01297 & 7.38443 & 4838 & 105 & 3.896 & 1.281 & $-$0.528 & 0.021 & 7 & 2 & 4741 & 97 & 2.899 & 0.832 & $-$0.515 & 0.143 & 9 & 12 & SCP \\
  100003-10388 & 38.84196 & 7.38046 & 5620 & 81 & 4.151 & 0.272 & $-$0.234 & 0.163 & 3 & 5 & 5556 & 122 & 4.128 & 0.160 & $-$0.283 & 0.141 & 2 & 5 & HQP \\
  100003-10399 & 38.15408 & 7.38761 & 5485 & 139 & 4.015 & 0.488 & $-$1.099 & 0.669 & 6 & 3 & 5398 & 276 & 3.534 & 0.790 & $-$1.004 & 0.584 & 4 & 64 & SCP \\
  100003-10420 & 38.86485 & 7.35709 & 6401 & 201 & 4.158 & 0.095 & $-$0.461 & 0.362 & 7 & 517 & 6481 & 213 & 4.152 & 0.104 & $-$0.481 & 0.370 & 8 & 139 & SCP \\
  100003-10430 & 38.50724 & 7.38084 & 5871 & 87 & 4.205 & 0.143 & $-$0.519 & 0.140 & 3 & 41 & 5849 & 81 & 4.191 & 0.156 & $-$0.604 & 0.156 & 2 & 25 & HQP \\
  100003-10466 & 38.15998 & 7.39121 & 3888 & 142 & 4.614 & 0.153 & $-$0.295 & 0.258 & 9 & 108 & 4090 & 117 & 4.662 & 0.515 & $-$0.122 & 0.267 & 6 & 1 & SCP \\
  100003-10506 & 38.81393 & 7.39422 & 4719 & 70 & 4.772 & 0.103 & $-$0.450 & 0.425 & 8 & 2 & 4771 & 112 & 3.980 & 0.918 & $-$0.316 & 0.198 & 10 & 463 & SCP \\
  100003-10515 & 38.51547 & 7.38997 & 5690 & 111 & 4.210 & 0.223 & $-$0.427 & 0.182 & 3 & 12 & 5675 & 142 & 4.171 & 0.210 & $-$0.454 & 0.205 & 2 & 9 & HQP \\
  100003-10519 & 37.91129 & 7.39526 & 5776 & 445 & 3.867 & 0.350 & $-$1.454 & 0.237 & 9 & 2 & -- & -- & -- & -- & -- & -- & -- & -- & SCP \\
  100003-10523 & 38.64669 & 7.39100 & 4967 & 98 & 4.484 & 0.236 & 0.050 & 0.133 & 5 & 1 & 4889 & 85 & 3.462 & 0.334 & $-$0.133 & 0.024 & 7 & 3 & HQP \\
  100003-10541 & 37.86663 & 7.38977 & 3763 & 19 & 4.764 & 0.091 & $-$0.064 & 0.132 & 6 & 2 & 3732 & 58 & 4.690 & 0.090 & $-$0.025 & 0.089 & 3 & 5 & HQP \\
  100003-10556 & 38.31917 & 7.38267 & 4877 & 112 & 4.267 & 0.551 & $-$0.086 & 0.154 & 7 & 156 & 4876 & 86 & 4.299 & 0.503 & 0.020 & 0.109 & 9 & 20 & SCP \\
  100003-10559 & 38.62032 & 7.39626 & 5232 & 113 & 4.218 & 0.479 & $-$0.175 & 0.194 & 7 & 113 & 5255 & 103 & 4.420 & 0.337 & $-$0.190 & 0.161 & 9 & 99 & HQP \\
  100003-10564 & 38.22547 & 7.38638 & 3933 & 90 & 4.627 & 0.111 & $-$0.243 & 0.429 & 7 & 14 & 3934 & 91 & 4.601 & 0.135 & $-$0.275 & 0.298 & 4 & 261 & HQP \\
  100003-10566 & 38.48288 & 7.39844 & 4887 & 88 & 4.214 & 0.780 & $-$0.391 & 0.110 & 6 & 4 & 4970 & 60 & 3.774 & 0.594 & $-$0.410 & 0.059 & 8 & 3 & SCP \\
  100003-10603 & 38.25412 & 7.40170 & 5272 & 287 & 3.156 & 0.513 & $-$1.455 & 0.567 & 9 & 12 & 5173 & 122 & 2.836 & 0.530 & $-$1.582 & 0.217 & 9 & 1 & SCP \\
  100003-10611 & 38.67809 & 7.39763 & 5155 & 113 & 4.600 & 0.255 & $-$0.363 & 0.164 & 6 & 1 & 5197 & 99 & 4.781 & 0.421 & $-$0.292 & 0.119 & 7 & 1 & HQP \\
  100003-1064 & 39.13597 & 6.79789 & 5462 & 153 & 3.620 & 0.409 & $-$1.286 & 0.228 & 9 & 1 & -- & -- & -- & -- & -- & -- & -- & -- & SCP \\
  ... & ... & ... & ... & ... & ... & ... & ... & ... & ... & ... & ... & ... & ... & ... & ... & ... & ... & ... & ... \\
  \hline
\end{tabular}
}
    \tablefoot{The columns are described in the text. The complete version of this table is available online on Vizier.}
    \end{sidewaystable*}

Table~\ref{statsSample} presents the statistics of the obtained stellar parameters using 105 colors, distributed according to their neighborhood radius and grouped according their photometric quality. The first column indicates the photometric quality subset. The second column shows the different neighborhood values, from 1 to 10. The third column indicates the number of stars with stellar parameters estimated according to their neighborhood radius. The remaining columns show the statistics (median, MAD, $\sigma$, minimum and maximum values) obtained for each stellar parameter.

    \begin{sidewaystable*}
    \caption{Statistics for the objects assigned to each neighborhood and according to their photometric quality (HQP or SCP).}
    \label{statsSample}
    \centering
    \begin{tabular}{lcc|ccccc|ccccc|ccccc}
     \hline
     \hline
     Quality & Neighbor- & N & \multicolumn{5}{c|}{\teff\ (K)} & \multicolumn{5}{c|}{\logg\ (dex)} & \multicolumn{5}{c}{\meta\ (dex)} \\
     sample & hood & & Median & MAD & $\sigma$ & Min. & Max. & Median & MAD & $\sigma$ & Min. & Max. & Median & MAD & $\sigma$ & Min. & Max. \\
     \hline              
HQP & 1 & 15491 & 5177 & 787 & 736 & 3186 & 11720 & 4.317 & 0.378 & 0.637 & 0.652 & 5.067 & $-$0.281 & 0.267 & 0.300 & $-$2.441 & 0.324 \\
 & 2 & 24302 & 6354 & 136 & 194 & 5210 & 8544 & 4.172 & 0.050 & 0.062 & 3.700 & 4.653 & $-$0.346 & 0.129 & 0.212 & $-$2.301 & 0.056 \\
 & 3 & 323418 & 5730 & 267 & 321 & 4560 & 11720 & 4.224 & 0.117 & 0.172 & 2.037 & 4.890 & $-$0.273 & 0.193 & 0.224 & $-$2.441 & 0.295 \\
 & 4 & 392880 & 5632 & 257 & 304 & 3669 & 11720 & 4.231 & 0.106 & 0.210 & 1.501 & 4.900 & $-$0.296 & 0.203 & 0.249 & $-$2.441 & 0.299 \\
 & 5 & 142215 & 5111 & 411 & 467 & 3391 & 15784 & 4.314 & 0.371 & 0.571 & 1.109 & 5.002 & $-$0.224 & 0.232 & 0.276 & $-$2.518 & 0.324 \\
 & 6 & 186845 & 5058 & 452 & 556 & 3267 & 9820 & 4.333 & 0.348 & 0.538 & 0.985 & 4.998 & $-$0.231 & 0.229 & 0.264 & $-$2.431 & 0.324 \\
 & 7 & 510792 & 4700 & 497 & 496 & 3253 & 14068 & 4.470 & 0.270 & 0.516 & 0.731 & 5.037 & $-$0.219 & 0.172 & 0.207 & $-$2.410 & 0.324 \\
 & 8 & 174394 & 3881 & 258 & 373 & 3186 & 11033 & 4.613 & 0.065 & 0.325 & 0.652 & 5.067 & $-$0.184 & 0.205 & 0.214 & $-$2.410 & 0.324 \\
 & 9 & 101152 & 3945 & 458 & 504 & 3186 & 11720 & 4.596 & 0.115 & 0.432 & 0.918 & 5.067 & $-$0.220 & 0.157 & 0.201 & $-$2.410 & 0.304 \\
 & 10 & 42106 & 4462 & 334 & 380 & 3237 & 11155 & 4.514 & 0.239 & 0.499 & 0.932 & 4.954 & $-$0.324 & 0.163 & 0.191 & $-$2.410 & 0.304 \\
SCP & 1 & 4 & 6364 & 85 & 1215 & 3958 & 6430 & 4.279 & 0.076 & 0.173 & 4.208 & 4.590 & $-$0.275 & 0.128 & 0.133 & $-$0.445 & $-$0.122 \\
 & 2 & 20457 & 6358 & 144 & 203 & 5309 & 8544 & 4.172 & 0.054 & 0.064 & 3.544 & 4.665 & $-$0.334 & 0.139 & 0.221 & $-$2.301 & 0.001 \\
 & 3 & 240658 & 5803 & 331 & 380 & 4560 & 11720 & 4.206 & 0.104 & 0.162 & 2.037 & 4.890 & $-$0.293 & 0.199 & 0.274 & $-$2.441 & 0.295 \\
 & 4 & 595710 & 5685 & 264 & 338 & 3606 & 11720 & 4.227 & 0.107 & 0.193 & 1.427 & 4.849 & $-$0.334 & 0.218 & 0.270 & $-$2.441 & 0.295 \\
 & 5 & 220049 & 5567 & 351 & 513 & 3466 & 17169 & 4.259 & 0.214 & 0.442 & 1.060 & 4.917 & $-$0.338 & 0.274 & 0.323 & $-$3.225 & 0.324 \\
 & 6 & 469090 & 5602 & 346 & 479 & 3314 & 16056 & 4.229 & 0.143 & 0.357 & 0.985 & 4.998 & $-$0.387 & 0.250 & 0.325 & $-$4.000 & 0.324 \\
 & 7 & 801559 & 5111 & 522 & 611 & 3267 & 16731 & 4.247 & 0.326 & 0.491 & 0.731 & 5.037 & $-$0.324 & 0.231 & 0.287 & $-$3.081 & 0.324 \\
 & 8 & 265713 & 4154 & 802 & 890 & 3186 & 16428 & 4.563 & 0.187 & 0.507 & 0.652 & 5.067 & $-$0.331 & 0.308 & 0.336 & $-$3.127 & 0.324 \\
 & 9 & 671631 & 3954 & 433 & 678 & 3186 & 14679 & 4.560 & 0.137 & 0.410 & 0.652 & 5.067 & $-$0.251 & 0.214 & 0.252 & $-$3.194 & 0.324 \\
 & 10 & 450020 & 4802 & 492 & 579 & 3186 & 14233 & 4.299 & 0.446 & 0.523 & 0.652 & 5.067 & $-$0.360 & 0.175 & 0.227 & $-$3.167 & 0.324 \\

\hline
    \end{tabular}
    \end{sidewaystable*}

Table~\ref{statsOthers} shows the differences ($\Delta$) obtained between our estimated stellar parameters and those from \citet{Wang2022} and \citet{Yang2024}, discriminated by the photometric quality samples and the different neighborhoods. The first column shows the two subsets with different photometric quality. The second column indicates the neighborhood values using 105 colors. The remaining columns report the statistics (median, MAD, $\sigma$) for each $\Delta$ value.

\begin{table*}
    \caption{Uncertainties in stellar parameter estimates between our method using 105 colors and other photometric estimates.}
    \label{statsOthers}
    \centering
    \begin{tabular}{lcccccccccc}
     \hline
     \hline
     Quality & Neighbor- & \multicolumn{3}{c}{$\Delta$\teff\ (K)} & \multicolumn{3}{c}{$\Delta$\logg\ (dex)} & \multicolumn{3}{c}{$\Delta$\meta\ (dex)} \\
     sample     &   hood & median & MAD  & $\sigma$ & median & MAD  & $\sigma$ & median & MAD  & $\sigma$ \\
     \hline              
\multicolumn{11}{l}{\sl Comparison with reference 1} \\
 HQP &  1 & $-$42 & 126 & 155 &  0.071 & 0.195 & 0.310 & 0.051 & 0.177 & 0.193 \\
 & 2 &  79 & 126 & 137 &  0.113 & 0.050 & 0.063 & 0.114 & 0.137 & 0.161 \\
 & 3 &   4 & 111 & 130 &  0.098 & 0.100 & 0.158 & 0.094 & 0.139 & 0.159 \\
 & 4 & $-$10 & 106 & 124 &  0.081 & 0.086 & 0.170 & 0.107 & 0.157 & 0.177 \\
 & 5 & $-$19 & 116 & 131 &  0.054 & 0.214 & 0.368 & 0.081 & 0.165 & 0.194 \\
 & 6 & $-$17 & 112 & 129 &  0.043 & 0.178 & 0.386 & 0.093 & 0.167 & 0.192 \\
 & 7 &  $-$9 &  86 &  95 &  0.003 & 0.166 & 0.372 & 0.107 & 0.148 & 0.177 \\
 & 8 & $-$13 &  85 &  94 &  0.059 & 0.088 & 0.282 & 0.222 & 0.225 & 0.233 \\
 & 9 &  $-$7 & 112 & 116 &  0.048 & 0.157 & 0.445 & 0.117 & 0.236 & 0.244 \\
& 10 & $-$14 &  91 &  98 &  0.029 & 0.150 & 0.382 & 0.060 & 0.143 & 0.163 \\

 SCP & 1 & 118 & 126 & 140 &  0.135 & 0.064 & 0.055 & 0.153 & 0.053 & 0.053 \\
 & 2 & 108 & 139 & 144 &  0.111 & 0.055 & 0.068 & 0.126 & 0.161 & 0.188 \\
 & 3 &  40 & 130 & 151 &  0.097 & 0.101 & 0.156 & 0.133 & 0.181 & 0.205 \\ 
 & 4 &  41 & 121 & 143 &  0.076 & 0.098 & 0.179 & 0.193 & 0.192 & 0.214 \\
 & 5 &  39 & 137 & 159 &  0.080 & 0.168 & 0.304 & 0.179 & 0.227 & 0.254 \\
 & 6 &  39 & 130 & 155 &  0.065 & 0.130 & 0.278 & 0.206 & 0.229 & 0.266 \\
 & 7 &   6 & 106 & 136 &  0.000 & 0.208 & 0.406 & 0.164 & 0.199 & 0.249 \\
 & 8 &  $-$8 & 132 & 230 &  0.075 & 0.160 & 0.436 & 0.192 & 0.265 & 0.305 \\
 & 9 & $-$19 & 129 & 279 &  0.040 & 0.236 & 0.547 & 0.179 & 0.242 & 0.270 \\
 & 10 & $-$27 & 112 & 154 & $-$0.119 & 0.429 & 0.526 & 0.177 & 0.196 & 0.220 \\
\multicolumn{11}{l}{\sl Comparison with reference 2} \\
HQP &  1 &  19 &  94 & 122 & $-$0.026 & 0.133 & 0.178 & $-$0.003 & 0.122 & 0.206 \\
 & 2 &  96 &  79 &  92 &  0.012 & 0.142 & 0.145 &  0.021 & 0.166 & 0.197 \\
 & 3 &  25 &  80 &  94 & $-$0.054 & 0.186 & 0.220 &  0.034 & 0.151 & 0.190 \\
 & 4 &  12 &  84 &  95 & $-$0.081 & 0.197 & 0.245 &  0.053 & 0.197 & 0.242 \\
 & 5 &  13 &  93 & 112 & $-$0.036 & 0.208 & 0.345 &  0.040 & 0.177 & 0.245 \\
 & 6 &  17 &  97 & 111 & $-$0.048 & 0.196 & 0.367 &  0.052 & 0.203 & 0.282 \\
 & 7 &  40 &  95 & 109 & $-$0.099 & 0.193 & 0.396 &  0.071 & 0.241 & 0.336 \\
 & 8 & 162 & 164 & 167 & $-$0.065 & 0.080 & 0.314 &  0.259 & 0.482 & 0.534 \\
 & 9 &  56 & 144 & 183 & $-$0.090 & 0.187 & 0.517 &  0.124 & 0.381 & 0.481 \\
 & 10 &  21 & 110 & 132 & $-$0.133 & 0.245 & 0.499 &  0.073 & 0.336 & 0.446 \\
SCP &  1 &  76 &  27 &  26 & $-$0.007 & 0.008 & 0.008 & $-$0.036 & 0.015 & 0.015 \\
 & 2 &  95 &  83 & 106 &  0.012 & 0.149 & 0.156 &  0.010 & 0.181 & 0.224 \\
 & 3 &  36 &  90 & 109 & $-$0.058 & 0.197 & 0.229 &  0.051 & 0.185 & 0.257 \\
 & 4 &  39 &  92 & 106 & $-$0.131 & 0.195 & 0.253 &  0.137 & 0.277 & 0.346 \\
 & 5 &  37 & 112 & 131 & $-$0.122 & 0.232 & 0.331 &  0.138 & 0.304 & 0.405 \\
 & 6 &  44 & 108 & 125 & $-$0.187 & 0.218 & 0.322 &  0.194 & 0.372 & 0.471 \\
 & 7 &  38 & 100 & 128 & $-$0.196 & 0.277 & 0.444 &  0.170 & 0.360 & 0.478 \\
 & 8 &  79 & 155 & 229 & $-$0.096 & 0.193 & 0.487 &  0.255 & 0.500 & 0.575 \\
 & 9 &  53 & 136 & 235 & $-$0.136 & 0.249 & 0.620 &  0.243 & 0.467 & 0.562 \\
 & 10 &  33 & 115 & 166 & $-$0.305 & 0.430 & 0.568 &  0.231 & 0.478 & 0.579 \\ 

\hline
    \end{tabular}
        \tablefoot{Statistics of the differences between our derived stellar parameters using 105 colors and those from \citet{Wang2022} and \citet{Yang2024} (comparison dataset), separated by the quality of their photometry and the proposed neighborhood values.}
\tablebib{
(1)~\citet{Wang2022};
(2)~\citet{Yang2024}.
}
\end{table*}

\section{Distribution of stellar parameters by neighborhood} 
\label{appendix:neighborhoods}

\begin{figure}
	\centering
    \includegraphics[width=0.95\columnwidth,trim=0.22cm 0 0 0,clip]{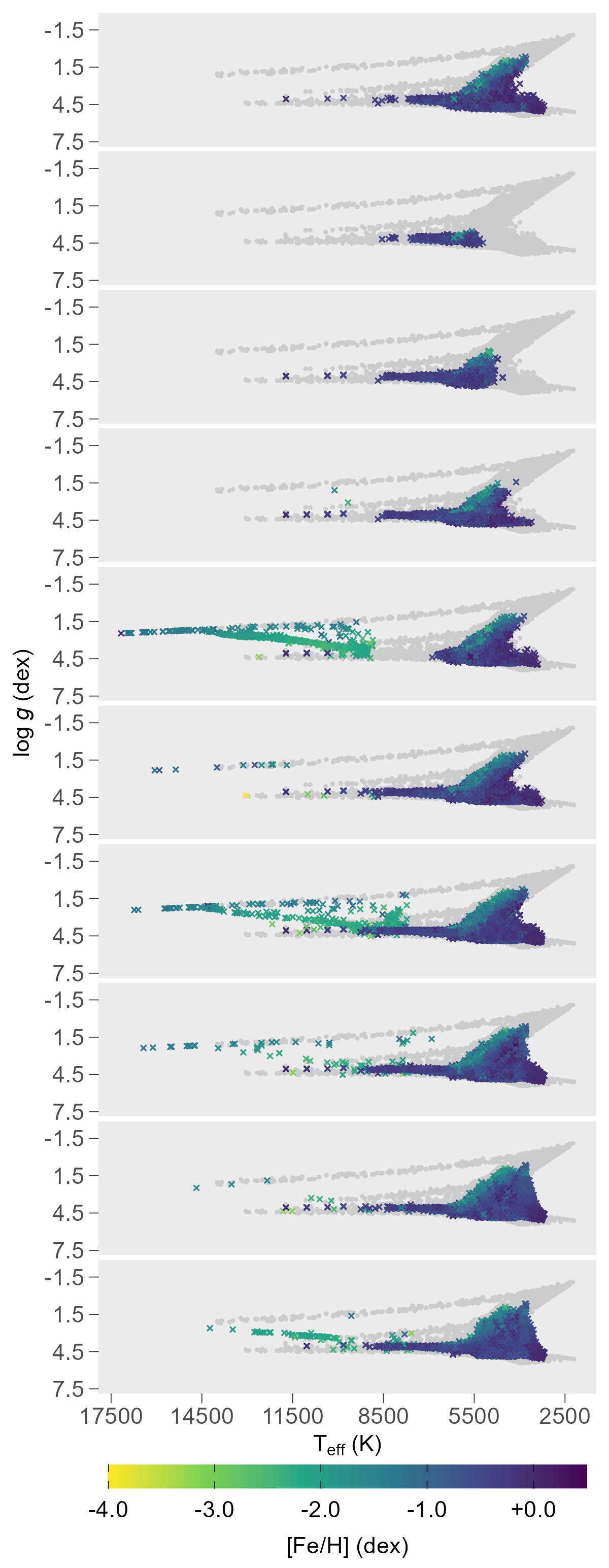}
	\caption{Kiel diagrams showing the distribution of the nearly 5.6 million stars in our sample, separated by neighborhood. Panels are ordered from 1 ({\sl top panel}) to 10 ({\sl bottom panel}). The gray symbols represent the reference dataset. The color bar uses the same scale as in Fig.~\ref{Kiel}, and \meta\ values range from $-$4.0 to +0.324\,dex.}
\label{Kiel2}
\end{figure}

Figure~\ref{Kiel2} presents the Kiel diagrams for the 5\,648\,486 stars for which stellar atmospheric parameters were derived using the complete set of 105 colors. The results are separated according to the neighborhood index assigned by the method, from neighborhood~1 in the upper panel to neighborhood~10 in the lower panel. The reference sample is shown in gray, while the estimated stellar parameters are color coded according to the inferred metallicity. This representation complements the statistical information provided in Sect.~\ref{AllColorsResults} and Table~\ref{statsSample} by illustrating how the distribution of the inferred parameters changes as progressively larger neighborhoods are required to identify suitable reference objects.

\section{Comparison of estimated atmospheric parameters}\label{AllComparations}

\begin{figure*}[!ht]
	\centering
        \includegraphics[width=0.95\textwidth]{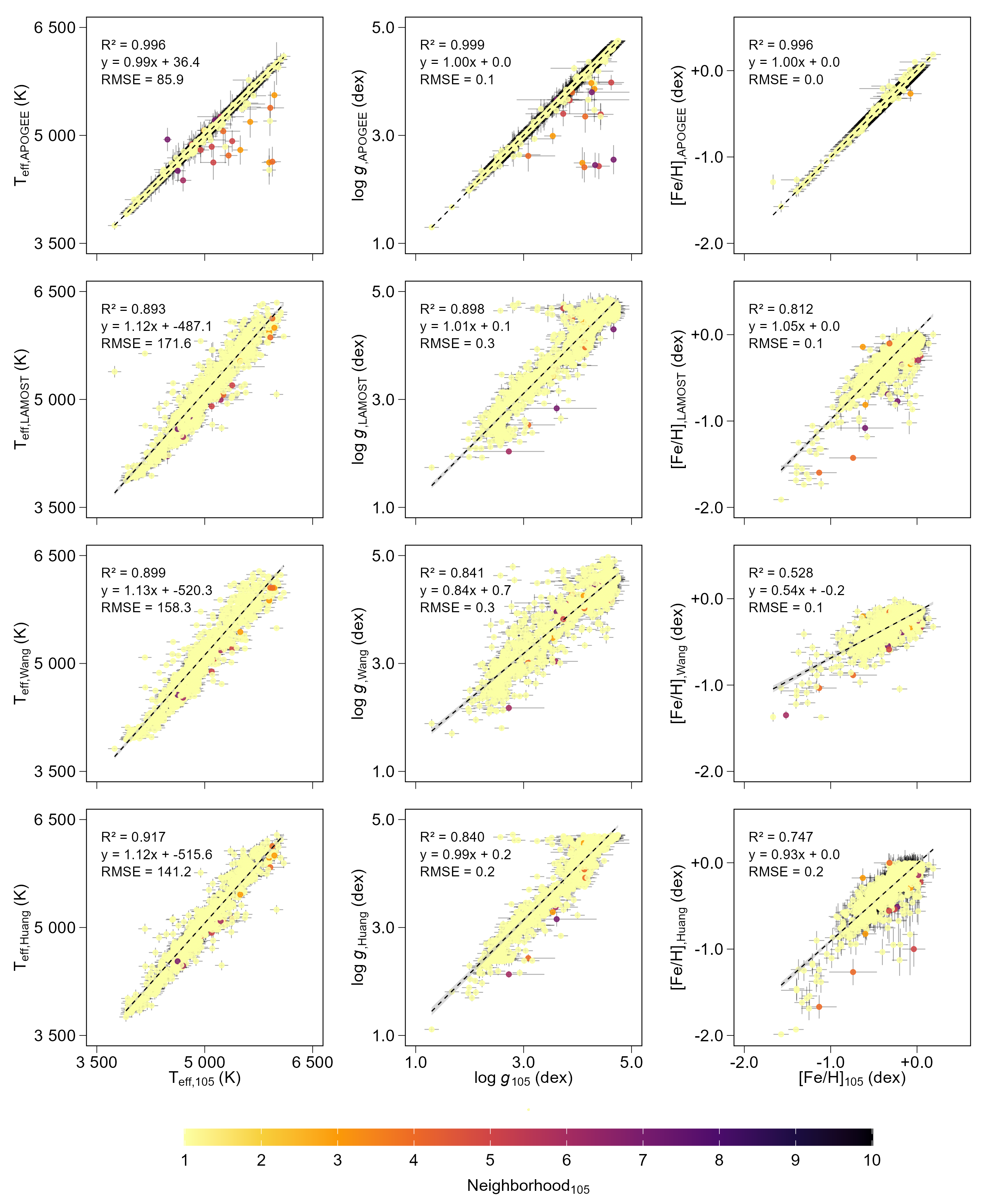}
	\caption{Comparison of the stellar parameter estimated by our method using 105 colors (x-axis) with those obtained from other sources (y-axis): APOGEE (first row), LAMOST (second row), \citet{Wang2022} (third row), and \citet{Yang2024} (fourth row). The color bar indicates the neighborhood index. The dashed line shows the best-fit linear regression, and the shaded region represents the 95\% confidence interval of the fit. The regression equation, coefficient of determination ($R^2$) and the root mean square error (RMSE) are indicated in each panel. Each point includes the error values reported by the sources indicated on each axis.}
	\label{CostrastALLAUTHORS}
\end{figure*}

Figure~\ref{CostrastALLAUTHORS} shows the comparison of stellar atmospheric parameters (\teff, \logg, and \meta) derived in this work using the 105-color method and those reported in the literature for 1\,236 stars in common among all surveys. The comparison includes spectroscopic parameters from APOGEE \citep{2022AJ....163..152S} and LAMOST DR9 \citep{bai2021first}, as well as photometric estimates from \citet{Wang2022} and \citet{Yang2024} based on J-PLUS DR1 and DR3 data, respectively.
A linear regression was applied to each panel as a diagnostic tool to assess the agreement between our estimates and the reference values. The dashed line represents the linear relationship of best fit, while the shaded region corresponds to the 95\% confidence interval of the fit. The regression equation, coefficient of determination ($R^2$), and root mean square error (RMSE) are presented for each panel.

\section{Correlation matrix between colors and stellar parameters}\label{MatrixCoor}

Matrix of correlations with colors between the parameters Temperature, Gravity and Metallicity and the 105 colors extracted from the information collected from J-PLUS and 2MASS survey. On one side, there is a colored bar associated with the correlation index for each of the cases. The order of the variables in the array viewed from top to bottom and left to right is: \teff , \logg , \meta, $u-J0378$, $u-J0395$, $u-J0410$, $u-J0430$, $u-g$, $u-J0515$, $u-r$, $u-J0660$, $u-i$, $u-J0861$, $u-z$, $u-J$, $u-H$, $u-K$, $J0378-J0395$, $J0378-J0410$, $J0378-J0430$, $J0378-g$, $J0378-J0515$, $J0378-r$, $J0378-J0660$, $J0378-i$, $J0378-J0861$, $J0378-z$, $J0378-J$, $J0378-H$, $J0378-K$, $J0395-J0410$, $J0395-J0430$, $J0395-g$, $J0395-J0515$, $J0395-r$, $J0395-J0660$, $J0395-i$, $J0395-J0861$, $J0395-z$, $J0395-J$, $J0395-H$, $J0395-K$, $J0410-J0430$, $J0410-g$, $J0410-J0515$, $J0410-r$, $J0410-J0660$, $J0410-i$, $J0410-J0861$, $J0410-z$, $J0410-J$, $J0410-H$, $J0410-K$, $J0430-g$, $J0430-J0515$, $J0430-r$, $J0430-J0660$, $J0430-i$, $J0430-J0861$, $J0430-z$, $J0430-J$, $J0430-H$, $J0430-K$, $g-J0515$, $g-r$, $g-J0660$, $g-i$, $g-J0861$, $g-z$, $g-J$, $g-H$, $g-K$, $J0515-r$, $J0515-J0660$, $J0515-i$, $J0515-J0861$, $J0515-z$, $J0515-J$, $J0515-H$, $J0515-K$, $r-J0660$, $r-i$, $r-J0861$, $r-z$, $r-J$, $r-H$, $r-K$, $J0660-i$, $J0660-J0861$, $J0660-z$, $J0660-J$, $J0660-H$, $J0660-K$, $i-J0861$, $i-z$, $i-J$, $i-H$, $i-K$, $J0861-z$, $J0861-J$, $J0861-H$, $J0861-K$, $z-J$, $z-H$, $z-K$, $J-H$, $J-K$, $H-K$.

\begin{figure*}
	\centering
        \includegraphics[width=1\textwidth]{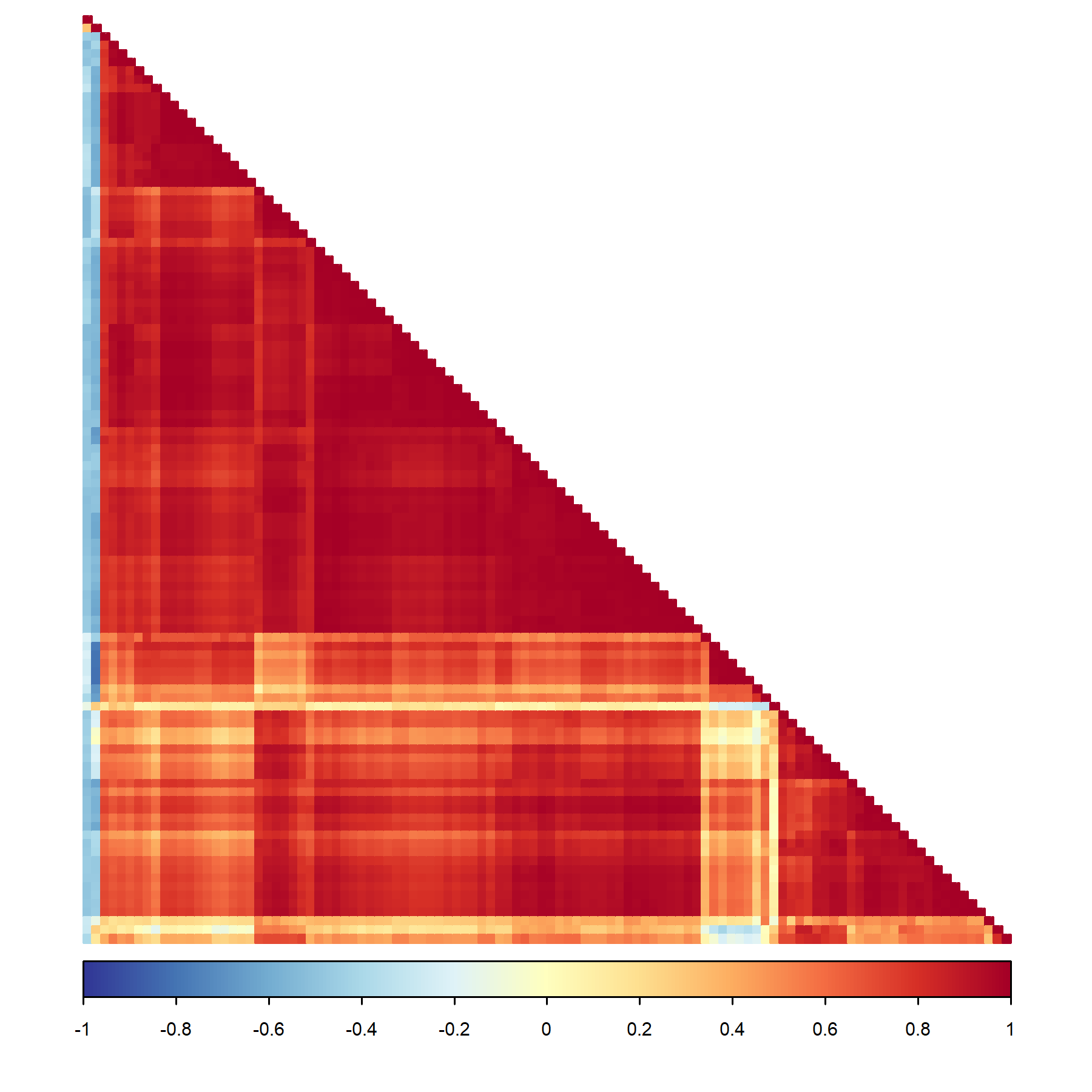}
	\caption{Matrix of correlation of parameters \teff, \logg\ and \meta, and the 105 colors obtained from the J-PLUS DR3 and 2MASS catalogs combined. Each row and column (105 in total) represent a color, order by their wavelengths, as indicated in Appendix~\ref{MatrixCoor}. The color bar indicates the different correlation values obtained.}
	\label{Coor}
\end{figure*}

\end{appendix}

\end{document}